\documentclass[12pt,draftcls, onecolumn]{IEEEtran}

\usepackage{amsmath}
\usepackage{amssymb}
\usepackage[dvips]{graphicx}
\usepackage{setspace}
\usepackage{epsfig}
\usepackage{amsmath}
\usepackage{amssymb}
\usepackage[dvips]{graphicx}
\usepackage{epsfig}

\newcommand{\E}{\mathbb{E}}

\newcommand{\figsize}{0.6}
\newcommand{\tr}{\mathbf{tr}}
\newcommand{\K}{\mathbf{K}}
\newcommand{\I}{\mathbf{I}}

\newcommand{\U}{\mathbf{U}}
\newcommand{\bLambda}{\mathbf{\Lambda}}
\newcommand{\dd}{\dagger}
\newcommand{\snr}{\textsc{snr}}

\newtheorem{theo}{Theorem}

\newtheorem{rem}{Remark}

\begin{document}

\title{On the Throughput and Energy Efficiency of Cognitive MIMO Transmissions}

\author{\authorblockN{Sami Akin and
M. Cenk Gursoy}\\
\thanks{S. Akin is with the Institute of Communications Technology, Leibniz Universit\"{a}t Hannover, 30167 Hanover, Germany. M. C. Gursoy is with the Department of Electrical Engineering and Computer Science, Syracuse University, Syracuse, NY, 13244.
(e-mail: sami.akin@ikt.uni-hannover.de, mcgursoy@syr.edu).}
\thanks{This work was supported by the National Science Foundation under Grants CCF -- 0546384 (CAREER)
and CCF -- 0917265. The material in
this paper was presented in part at the IEEE Wireless Communications and Networking Conference (WCNC) in March 2011. }}
\date{}

\maketitle
\begin{spacing}{1.8}
\vspace{-.5cm}
\begin{abstract}
In this paper, throughput and energy efficiency of cognitive multiple-input multiple-output (MIMO) systems operating under quality-of-service (QoS) constraints, interference limitations, and imperfect channel sensing, are studied. It is assumed that transmission power and covariance of the input signal vectors are varied depending on the sensed activities of primary users (PUs) in the system. Interference constraints are applied on the transmission power levels of cognitive radios (CRs) to provide protection for the PUs whose activities are modeled as a Markov chain. Considering the reliability of the transmissions and channel sensing results, a state-transition model is provided.  Throughput is determined by formulating the effective capacity. First derivative of the effective capacity is derived in the low-power regime and the minimum bit energy requirements in the presence of QoS limitations and imperfect sensing results are identified. Minimum energy per bit is shown to be achieved by beamforming in the maximal-eigenvalue eigenspace of certain matrices related to the channel matrix. In a special case, wideband slope is determined for more refined analysis of energy efficiency. Numerical results are provided for the throughput for various levels of buffer constraints and different number of transmit and receive antennas. The impact of interference constraints and benefits of multiple-antenna transmissions are determined. It is shown that increasing the number of antennas when the interference power constraint is stringent is generally beneficial. On the other hand, it is shown that under relatively loose interference constraints, increasing the number of antennas beyond a certain level does not lead to much increase in the throughput.
\end{abstract}

\end{spacing}

\begin{spacing}{1.8}

\section{Introduction}

Cognitive Radio (CR), which has emerged as a method to tackle the spectrum scarcity and variability in both time and space, calls for dynamic access strategies that adapt to the electromagnetic environment \cite{barbarossa}. Performance of cognitive radio systems has been studied extensively in recent years, and a detailed description of different CR models and an overview of recent approaches can be found in \cite{akyildiz}, \cite{Goldsmith2} and \cite{Zhao}. For instance, three different paradigms, namely underlay, overlay and interweave operation of cognitive radio systems, were discussed in \cite{Goldsmith2}. In underlay CR networks, cognitive secondary users (SUs) can coexist with the primary users (PUs) and transmit concurrently as long as they adhere to strict limitations on the interference inflicted on the PUs. This model is also known as spectrum sharing. On the other hand, in interweave CR networks,  SUs initially perform channel sensing and opportunistically access only the spectrum holes in which the primary users are inactive. These two methods of spectrum sharing and opportunistic spectrum access can also be combined for improved performance. For instance, Kang \textit{et al.} in \cite{Zhang} analyzed a hybrid model in which SUs first sense the frequency bands and detect the PU activity. Subsequently, cognitive radio transmission is performed at two different power levels depending on the sensed PU activity. More specifically, if the PUs are sensed to be active, secondary transmission still occurs but with reduced power level in order to lower the interference within tolerable levels. In such modes of cognitive operation, sensing the activities of PUs is a critical issue that has been studied and analyzed extensively (see e.g., \cite{Zhao2}, \cite{Zhao3}) since the inception of the CR concept.

Another advancement in communications technology is multiple-antenna communications. It is well-known that employing multiple antennas at the receiver and transmitter ends of a communication system can improve the performance levels by providing significant gains in the throughput and/or reliability of transmissions. Therefore, there has been much interest in understanding and analyzing multiple-input multiple-output (MIMO) channels and numerous comprehensive studies have been conducted \cite{goldsmith}, \cite{telatar}. In most studies, ergodic Shannon capacity formulations are considered as the performance metrics \cite{Lozano}, \cite{Lozano2}, \cite{Sandhu}. For instance, the authors in \cite{Lozano} and \cite{Lozano2} studied multiple-antenna ergodic channel capacity and provided analytical characterizations of the impact of certain factors such as antenna correlation, co-channel interference, Ricean factors, and polarization diversity. It should be noted that ergodic capacity generally does not take into account any delay, buffer, or queueing constraints at the transmitter.

In \cite{Gursoy_IT}, the throughput of MIMO systems in the presence of statistical queuing constraints was investigated. Effective capacity was employed as the metric to measure the performance under quality-of-service (QoS) constraints. Effective capacity characterizes the maximum constant arrival rate that can be supported by a system under statistical limitations on buffer violations \cite{Wu}. There have been several studies on effective capacity in various communication settings \cite{Tang}, \cite{Tang2}. Recently, the authors in \cite{Mittel} considered the maximization of effective capacity in a single-user multi-antenna system with covariance knowledge, and the authors in \cite{Liu} studied the effective capacity of a class of multiple-antenna wireless systems subject to Rayleigh flat fading.

Recently, cognitive MIMO radio models have also been considered since having multiple antennas can provide higher performance levels for the SUs and lead to better protection of PUs. Modeling a channel setting with a single licensed user and a single cognitive user, that is equivalent to an interference channel with degraded message sets, the authors in \cite{sridharan} focused on the fundamental performance limits of a cognitive MIMO radio network, and they showed that under certain conditions, the achievable region is optimal for a portion of the capacity region that includes the sum capacity. In \cite{YingJun}, three scenarios, namely when the secondary transmitter (ST) has complete, partial, or no knowledge about the channels to the primary receivers (PRs), was considered, and maximization of the throughput of the SU, while keeping the interference temperature at the PRs below a certain threshold, was investigated. Furthermore, in \cite{feifei}, the authors proposed a practical CR transmission strategy consisting of three major stages, namely, environment learning that applies blind algorithms to estimate the spaces that are orthogonal to the channels from the PR, channel training that uses training signals and employs the linear-minimum-mean-square-error (LMMSE)-based estimator to estimate the effective channel, and data transmission. Considering imperfect estimations in both learning and training stages, they derived a lower bound on the ergodic capacity that is achievable by the CR in the data-transmission stage. In another study \cite{RuiZhang}, the authors proposed a practical cognitive beamforming scheme that does not require any prior knowledge of the CR-PR channels, but exploits the time-division-duplex operation mode of the PR link and the channel reciprocities between CR and PR terminals, utilizing an idea called effective interference channel, that is estimated at the CR terminal via periodically observing the PR transmissions. It was also shown in \cite{Samir} that the asymptotes of the achievable transmission rates of the opportunistic (secondary) link are obtained in the regime of large numbers of antennas. Another study of cognitive MIMO radios was conducted in \cite{Zhang-Liang}.

The above-mentioned references have not addressed considerations related to energy efficiency and QoS provisioning in cognitive MIMO channels. In our prior work, we studied the impact of QoS requirements in single-antenna cognitive radio systems. In particular, we considered a CR model in which SUs transmit with two different transmission rates and power levels depending on the activities of PUs under QoS
constraints. In \cite{paper1}, the ST senses only one channel and then depending on the channel sensing results, it chooses its transmission policy, whereas in \cite{paper2} the ST senses more than one channel and chooses the best channel for transmission under interference power limits and QoS constraints. In \cite{paper3}, effective capacity limits of a CR model is analyzed with imperfect channel side information (CSI) at the transmitter and the receiver.

In this article, we focus on a cognitive MIMO system operating under QoS constraints. In particular, we investigate the achievable throughput levels and also study the performance in the low-power regime in order to address the energy efficiency. We analyze the impact of imperfect sensing results and interference limitations on the performance, and determine energy-efficient transmission strategies in the low-power regime. In the system model, we consider two different transmission policies depending on the activities of PUs and interference power threshold required to protect the PUs. Essentially, we have a hybrid, sensing-based spectrum sharing model of cognitive radio operation as described in \cite{Zhang}. We consider a general cognitive MIMO link where fading coefficients have arbitrary distributions and are possibly correlated across antennas. Moreover, we model the received interference signals from the primary transmitters correlated as well. We assume that the ST and secondary receiver (SR) have perfect side information regarding their own channels. The contributions of the paper can be summarized as follows:
\begin{enumerate}
  \item We identify a joint state-transition model, considering the reliability of the transmissions and taking into account the channel sensing decisions and their correctness.
  \item We provide a formulation of the throughput metric (effective capacity) in terms of transmission rates and state transition probabilities which depend on sensing reliability and primary user activity.
  \item We obtain expressions for the first and second derivative of the effective capacity at $\textsc{snr}=0$, and determine the minimum energy per bit in the presence of QoS limitations and imperfect sensing results.
\end{enumerate}

The organization of the paper is as follows. We provide the cognitive MIMO radio model and describe the transmission power and interference constraints in Section \ref{channel model}. In Section \ref{state transition}, we construct a state transition model for CR transmission and identify the throughput under QoS constraints, and show the relation between the effective capacity and ergodic capacity. Finding the first and second derivatives of effective capacity at $\textsc{snr}=0$, we analyze in Section \ref{effective capacity low power} the energy efficiency in the low-power regime. In Section \ref{Numeric}, we provide numerical results. We conclude in Section \ref{Conclusion}. Proofs are relegated to the Appendix.

\section{Channel Model, Power Constraints, and Input Covariance}\label{channel model}

\subsection{Channel Model}
\begin{figure}
\begin{center}
\includegraphics[width =0.7\textwidth]{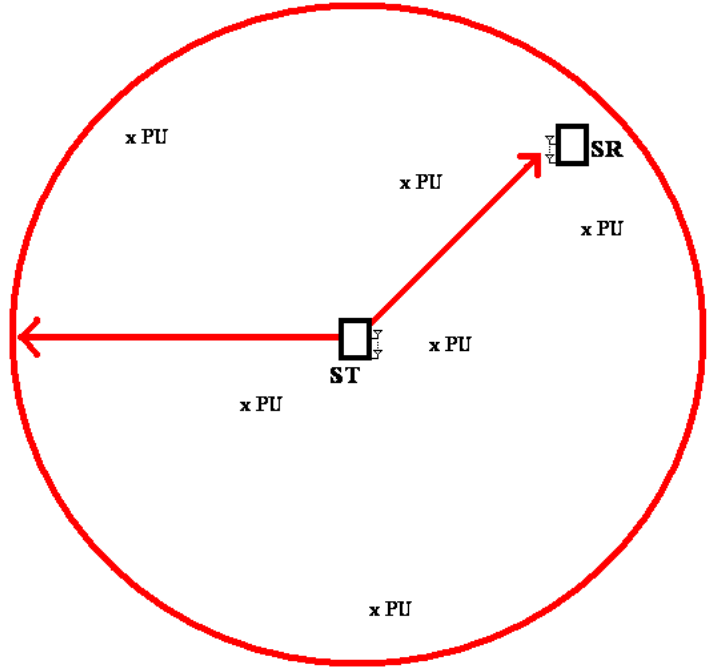}
\caption{The cognitive radio channel model.}
\label{fig:resimMIMO}
\end{center}
\end{figure}

As seen in Figure \ref{fig:resimMIMO}, we consider a setting in which a single ST communicates with a single SR in the presence of possibly multiple PUs. We consider a cognitive MIMO radio model and assume that the ST and SR are equipped with $M$ and $N$ antennas, respectively. In a flat fading channel, we can express the channel input-output relation as
\begin{equation}\label{input-output 1}
\textbf{y}=\textbf{H}\mathbf{x}+\mathbf{n}+\mathbf{s}
\end{equation}
if the PUs are active in the channel, and as
\begin{equation}\label{input-output 2}
\mathbf{y}=\mathbf{H}\mathbf{x}+\mathbf{n}
\end{equation}
if the PUs are absent. Above, $\mathbf{x}$ denotes the $M\times1-$dimensional transmitted signal vector of ST, and $\mathbf{y}$ denotes the $N\times1-$dimensional received signal vector at the SR. In ($\ref{input-output 1}$) and ($\ref{input-output 2}$), $\mathbf{n}$ is an $N\times1-$dimensional zero-mean Gaussian random vector with a covariance matrix
$\mathbb{E}\{\mathbf{n}\mathbf{n}^{\dag}\}=\sigma_{n}^{2}\mathbf{I}$ where $\mathbf{I}$ is the
identity matrix. In ($\ref{input-output 1}$), $\mathbf{s}$ is an $N\times1-$dimensional vector of the
sum of active PUs' faded signals arriving at the secondary receiver. Considering that the
vector $\mathbf{s}$ can have correlated components, we express its covariance matrix as
$\mathbb{E}\{\mathbf{s}\mathbf{s}^{\dag}\}=N\sigma_{s}^{2} \mathbf{K}_{s}$ where $\sigma_s^2$ is the
variance of each component of $\mathbf{s}$ and $\tr(\mathbf{K}_{s})=1$. Finally, in
($\ref{input-output 1}$) and ($\ref{input-output 2}$), $\mathbf{H}$ denotes the $N\times M$
dimensional random channel matrix whose components are the fading coefficients between the
corresponding antennas at the secondary transmitting and receiving ends. We consider a block-fading
scenario and assume that the realization of the matrix $\mathbf{H}$ remains fixed over a block
duration of $T$ seconds and changes independently from one block to another.

\subsection{Power and Interference Constraints}
We assume that the SUs initially perform channel sensing to detect the activities of PUs, and then depending on the channel sensing results, they choose the transmission
strategy. More specifically, if the channel is sensed as busy, the transmitted signal vector is $\mathbf{x}_1$. Otherwise, the signal is $\mathbf{x}_2$. When the channel is sensed as busy, the average energy of the channel input is
\begin{align}\label{power constraint 1}
\mathbb{E}\{||\mathbf{x}_{1}||^{2}\}&=\frac{P_{1}}{B}.
\end{align}
On the other hand, if the channel is detected to be idle, the average energy becomes
\begin{equation}\label{power constraint 2}
\mathbb{E}\{||\mathbf{x}_{2}||^{2}\} = \frac{P_{2}}{B}.
\end{equation}
In ($\ref{power constraint 1}$) and ($\ref{power constraint 2}$), $B$ is the bandwidth of the system. Note that under the assumption that $B$ complex input vectors are transmitted every second, the above energy levels imply that the transmission powers are $P_1$ and $P_2$, depending on the sensing results.

We first note that $P_1$ and $P_2$ are upper bounded by $P_{\max}$, which represents the maximum transmission power capabilities of cognitive transmitters. In a cognitive radio setting, transmission power levels are generally further restricted in order to limit the interference inflicted on the PUs. As a first measure, we assume that $P_{1}=\mu P_{2}$ where $0 \leq \mu \leq1$. Hence, smaller transmission power is used when the channel is sensed as busy, and we basically have
\begin{gather}
P_1 \le P_2 \le P_{\max}.
\end{gather}
Additionally, we consider a practical scenario in which errors such as miss-detections and false-alarms possibly
occur in channel sensing. We denote the correct-detection and false-alarm probabilities by $P_d$ and
$P_f$, respectively. We note the following two cases. When PUs are active and this
activity is sensed correctly (which happens with probability $P_d$ or equivalently $P_d$ fraction of the time on the average), then SUs
transmit with average power $P_1$. On the other hand, if the PU activity is missed in
sensing (which occurs with probability $1-P_d$), SUs send the information with
average power $P_2$. In both cases, PUs experience interference proportional to the product of the transmission power, average fading power, and path loss in the channel between the ST and PUs. In order to limit the
average interference, we impose the following constraint
\begin{equation}\label{Power Threshold}
P_{d}P_{1}+(1-P_{d})P_{2}\leq P_{int}
\end{equation}
where $P_{int}$ can be seen as the average interference constraint normalized by the average fading power and path loss\footnote{For instance, if average transmission power is limited by $P_{int}$ when the primary users are active, the average interference experienced at a given primary receiver will be limited by $P_{int} \frac{c}{d^{\varsigma}} \E\{z\}$ where $z$ is the magnitude square of the fading in the channel between the secondary transmitter and primary receiver, $d$ is the distance between them, $\varsigma$ is the path loss exponent, and $c$ is some constant related to the path loss model.}.
We note that a similar formulation for the average interference constraint was considered in \cite{Zhang}.
Noting the assumption that $P_{1}=\mu P_{2}$ for some $\mu \in [0,1]$, we can
rewrite ($\ref{Power Threshold}$) as
\begin{equation}\label{Power Threshold_new}
P_{d}\mu P_{2}+(1-P_{d})P_{2}\leq P_{int},
\end{equation}
which implies that
$
P_2 \le \frac{P_{int}}{P_d \mu + (1-P_d)}.
$
Considering the maximum of the average power, we can write
\begin{gather} \label{eq:P2upperbound}
P_2 \le \min\left\{P_{\max}, \frac{P_{int}}{P_d \mu + (1-P_d)}\right\}.
\end{gather}
Note that for given $\mu$ and detection probability $P_d$, if the interference constraints are relatively relaxed and we have $\frac{P_{int}}{P_d \mu + (1-P_d)} \ge P_{\max}$, then we can choose to operate at $P_2 = P_{\max}$ and $P_1 = \mu P_{\max}$. Otherwise, interference constraints will dictate the transmission power levels.

From (\ref{Power Threshold}), we can also, for given $P_2$, $P_{int}$ and $P_d$, obtain
\begin{gather}\label{Power Threshold_new_values}
\mu\leq \min\left\{\max\left\{\frac{P_{int}-P_{2}(1-P_{d})}{P_{2}P_{d}}, 0 \right\}, 1 \right\}.
\end{gather}
From above, we see that if $P_2 (1-P_d) \ge P_{int}$, then $\mu = 0$ and hence no transmission is performed by the ST when the channel is sensed as busy.


\begin{figure}
\begin{center}
\includegraphics[width =\figsize\textwidth]{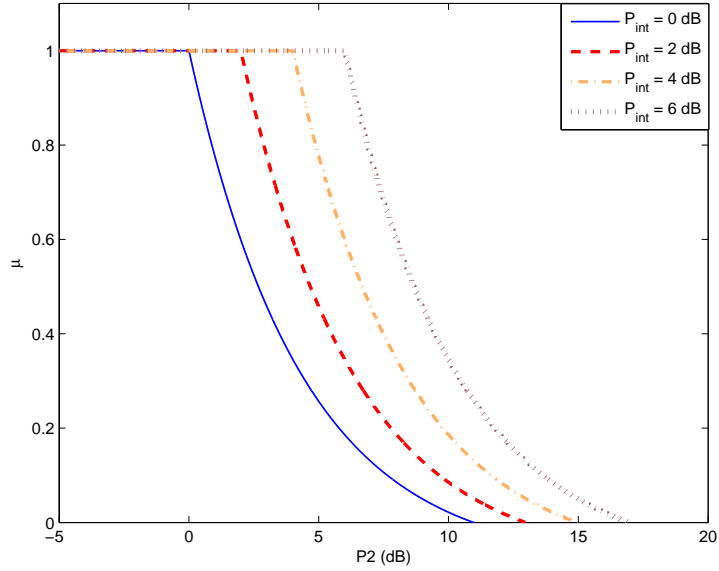}
\caption{The ratio $\mu = \frac{P_{1}}{P_2}$ vs. $P_{2}$.}
\label{fig:fig3}
\end{center}
\end{figure}

In order to illustrate some of the interactions between the parameters discussed above, we plot, in Fig. $\ref{fig:fig3}$, the ratio $\mu = \frac{P_1}{P_2}$ as a function of $P_{2}$, the power level adapted when the channel is sensed as idle, for different values of power interference constraints $P_{int}$. In all cases, we have $\mu = 1$ for small values of $P_{2}$, while $\mu$ diminishes to zero as $P_2$ increases due to the presence of interference constraints. Note also that we reach $\mu = 0$ at smaller values of $P_2$ under more stringent interference constraints.

\subsection{Input Covariance Matrix}
Finally, we note that in addition to having different levels of transmission power, directionality of the transmitted signal vectors might also be different depending on the channel
sensing results. We define the normalized input covariance matrix of $\mathbf{x}_1$ as
\begin{equation}\label{covariance 1}
\mathbf{K}_{x_{1}}=\frac{\mathbb{E}\{\mathbf{x}_{1} \mathbf{x}_{1}^{\dag}\}}{P_{1}/B}
\end{equation}
if the channel is busy, and that of $\mathbf{x}_2$ as
\begin{equation}\label{covariance 2}
\mathbf{K}_{x_{2}}=\frac{\mathbb{E}\{\mathbf{x}_{2} \mathbf{x}_{2}^{\dag}\}}{P_{2}/B}
\end{equation}
if the channel is idle. Note that the traces of normalized covariance matrices are
\begin{equation}\label{bound 1}
\tr(\mathbf{K}_{x_{1}}) = 1
\end{equation}
and
\begin{equation}\label{bound 2}
\tr(\mathbf{K}_{x_{2}}) = 1.
\end{equation}


\section{State Transition Model and Channel Throughput}\label{state transition}
\subsection{State Transition Model}
Depending on
channel sensing results and their correctness, we have four scenarios:
\begin{enumerate}
  \item Channel is busy, and is detected as busy (correct detection),
  \item Channel is busy, but is detected as idle (miss-detection),
  \item Channel is idle, but is detected as busy (false alarm),
  \item Channel is idle, and is detected as idle (correct detection).
\end{enumerate}

Using the notation $\mathbb{E}\{(\mathbf{s}+\mathbf{n})(\mathbf{s}+
\mathbf{n})^{\dag}\}=\mathbb{E}\{\mathbf{s}\mathbf{s}^{\dag}\} +\mathbb{E}\{\mathbf{n}
\mathbf{n}^{\dag}\}=N\sigma_s^2\K_s + \sigma_n^2 \I = \sigma_{n}^{2}\mathbf{K}_{z}$ where $\tr(\mathbf{K}_{z})=\frac{N(\sigma_{s}^{2}+
\sigma_{n}^{2})}{\sigma_{n}^{2}}$, we can express the instantaneous channel capacities in the above
four scenarios as follows:
\begin{align}\label{capacity}
&C_{1}=B\max_{\substack{\mathbf{K}_{x_{1}}\succeq0\\\tr(\mathbf{K}_{x_{1}}) = 1}}\log_{2}\det\left[ \mathbf{I}+\frac{\mu P_{2}}{B\sigma_{n}^{2}}\mathbf{H}\mathbf{K}_{x_{1}}\mathbf{H}^{\dag}\mathbf{K}_{z}^{-1}\right] =B\max_{\substack{\mathbf{K}_{x_{1}}\succeq0\\\tr(\mathbf{K}_{x_{1}})=1}}\log_{2}\det\left[\mathbf{I}+\mu N\textsc{snr}\mathbf{H}\mathbf{K}_{x_{1}}\mathbf{H}^{\dag}\mathbf{K}_{z}^{-1}\right],\nonumber\\
&C_{2}=B\max_{\substack{\mathbf{K}_{x_{2}}\succeq0\\\tr(\mathbf{K}_{x_{2}})=1}}\log_{2}\det\left[\mathbf{I}+\frac{P_{2} }{B\sigma_{n}^{2}}\mathbf{H}\mathbf{K}_{x_{2}}\mathbf{H}^{\dag}\mathbf{K}_{z}^{-1}\right] =B\max_{\substack{\mathbf{K}_{x_{2}}\succeq0\\\tr(\mathbf{K}_{x_{2}})=1}}\log_{2}\det\left[\mathbf{I}+ N\textsc{snr}\mathbf{H}\mathbf{K}_{x_{2}}\mathbf{H}^{\dag}\mathbf{K}_{z}^{-1}\right],\nonumber\\
&C_{3}=B\max_{\substack{\mathbf{K}_{x_{1}}\succeq0\\\tr(\mathbf{K}_{x_{1}})=1}}\log_{2}\det\left[\mathbf{I}+\frac{\mu P_{2}}{B\sigma_{n}^{2}}\mathbf{H}\mathbf{K}_{x_{1}}\mathbf{H}^{\dag}\right] =B\max_{\substack{\mathbf{K}_{x_{1}}\succeq0\\\tr(\mathbf{K}_{x_{1}})=1}}\log_{2}\det\left[\mathbf{I}+\mu N\textsc{snr}\mathbf{H}\mathbf{K}_{x_{1}}\mathbf{H}^{\dag}\right],\nonumber\\
&C_{4}=B\max_{\substack{\mathbf{K}_{x_{2}}\succeq0\\\tr(\mathbf{K}_{x_{2}})=1}}\log_{2}\det\left[\mathbf{I}+\frac{P_{2} }{B\sigma_{n}^{2}}\mathbf{H}\mathbf{K}_{x_{2}}\mathbf{H}^{\dag}\right] =B\max_{\substack{\mathbf{K}_{x_{2}}\succeq0\\\tr(\mathbf{K}_{x_{2}})=1}}\log_{2}\det\left[\mathbf{I}+ N\textsc{snr}\mathbf{H}\mathbf{K}_{x_{2}}\mathbf{H}^{\dag}\right].
\end{align}
Above, we define $\textsc{snr}=\frac{\mathbb{E}\{||\mathbf{x}_{2}||^{2}\}}{\mathbb{E}\{||\mathbf{n}||^{2}\}}=\frac{P_{2}}{NB\sigma_{n}^{2}}$ as the signal-to-noise ratio when the channel is sensed as idle. If, on the other hand, the channel is sensed as busy, signal-to-noise ratio is $\mu \textsc{snr}$ since the transmission power is $P_1 = \mu P_2$. We also note that since $\mathbf{K}_{z}$ is a positive definite matrix and its eigenvalues are greater
than or equal to 1, $\mathbf{K}_{z}^{-1}$ is a positive definite matrix with eigenvalues
$\frac{\sigma_{n}^{2}}{N(\sigma_{n}^{2}+\sigma_{s}^{2})}\le\lambda_{i}\le 1$.

\begin{figure}
\begin{center}
\includegraphics[width =\figsize\textwidth]{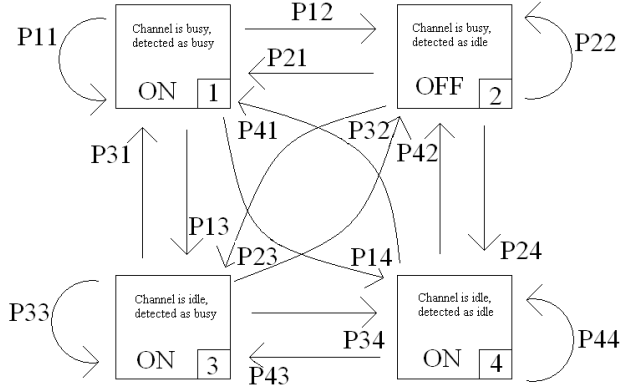}
\caption{State transition model for the cognitive radio channel. The numbered label for each state is given on the bottom-right corner of the box representing the state.} \label{fig:fig1}
\end{center}
\end{figure}

The secondary transmitter is assumed to send the data at two different rates depending on the sensing
results. If the channel is detected as busy as in scenarios 1 and 3, the transmission rate is
\begin{equation}\label{r1}
r_{1}=B\max_{\substack{\mathbf{K}_{x_{1}}\succeq0\\\tr(\mathbf{K}_{x_{1}})=1}}\log_{2}\det\left[\mathbf{I}+\mu N\textsc{snr}\mathbf{H}\mathbf{K}_{x_{1}}\mathbf{H}^{\dag}\mathbf{K}_{z}^{-1}\right],
\end{equation}
and if the channel is detected as idle as in scenarios 2 and 4, the transmission rate is
\begin{equation}\label{r2}
r_{2}=B\max_{\substack{\mathbf{K}_{x_{2}}\succeq0\\\tr(\mathbf{K}_{x_{2}})=1}}\log_{2}\det\left[\mathbf{I}+ N\textsc{snr}\mathbf{H}\mathbf{K}_{x_{2}}\mathbf{H}^{\dag}\right].
\end{equation}
In scenarios 1 and 4, sensing decisions are correct and transmission rates match the channel capacities, i.e., we have $r_1 = C_1$ in scenario 1, and $r_2 = C_4$ in scenario 4. In these cases, we assume that reliable communication is achieved. On the other hand, sensing errors in scenarios 2 and 3 lead to mismatches. We first establish the following result. Note that $\mathbf{H}\mathbf{K}_{x_{1}}\mathbf{H}^{\dag}$ and $\K_z^{-1}$ are 
are Hermitian matrices, they can be written as \cite[Theorem 4.1.5]{Matrix Analysis}
\begin{align} \label{eq:spectral}
\mathbf{H}\mathbf{K}_{x_{1}}\mathbf{H}^{\dag}  = \mathbf{A}  = \U_A \bLambda_A \U_A^\dd
\quad \text{and} \quad
\K_z^{-1} = \U_{K_z^{-1}} \bLambda_{K_z^{-1}} \U_{K_z^{-1}}^\dd
\end{align}
where $\U_A$ and $\U_{K_z^{-1}}$ are unitary matrices and $\bLambda_A$ and $\bLambda_{K_z^{-1}}$ are real diagonal matrices, consisting of the eigenvalues of $\mathbf{A }= \mathbf{H}\mathbf{K}_{x_{1}}\mathbf{H}^{\dag}$ and $\K_z^{-1}$, respectively.
Now, we can write
\begin{align}
\det\left[\mathbf{I}+\mu N\textsc{snr}\mathbf{H}\mathbf{K}_{x_{1}}\mathbf{H}^{\dag}\mathbf{K}_{z}^{-1}\right]&=\det\left[\mathbf{I}+\mu N\textsc{snr}\mathbf{A}\mathbf{K}_{z}^{-1}\right]\label{eq:det-inequality0}\\ &=\det\left[\left(
              \begin{array}{cc}
                \mathbf{U}_{A} & 0 \\
                0 & \mathbf{U}_{K_{z}^{-1}} \\
              \end{array}
            \right)\left(
                     \begin{array}{cc}
                       \mu N\textsc{snr}\mathbf{\Lambda}_{A} & -\mathbf{I} \\
                       \mathbf{I} & \mathbf{\Lambda}_{K_{z}^{-1}} \\
                     \end{array}
                   \right)\left(
                            \begin{array}{cc}
                              \mathbf{U}_{A}^{\dag} & 0 \\
                              0 & \mathbf{U}_{K_{z}^{-1}}^{\dag} \\
                            \end{array}
                          \right)
\right]\nonumber\\ &=\det\left[\mathbf{U}_{A}\mathbf{U}_{K_{z}^{-1}}\right]\det\left[\mathbf{I}+\mu N\textsc{snr}\mathbf{\Lambda}_{A}\mathbf{\Lambda}_{K_{z}^{-1}}\right]\det\left[\mathbf{U}_{A}^{\dag}\mathbf{U}_{K_{z}^{-1}}^{\dag}\right]\\
&\leq\det\left[\mathbf{U}_{A}\mathbf{U}_{K_{z}^{-1}}\right]\det\left[\mathbf{I}+\mu N\textsc{snr}\mathbf{\Lambda}_{A}\right]\det\left[\mathbf{U}_{A}^{\dag}\mathbf{U}_{K_{z}^{-1}}^{\dag}\right] \label{eq:det-inequality}\\
&=\det\left[\mathbf{I}+\mu N\textsc{snr}\mathbf{A}\right]=\det\left[\mathbf{I}+\mu N\textsc{snr}\mathbf{H}\mathbf{K}_{x_{1}}\mathbf{H}^{\dag}\right]\label{eq:det-inequality1}.
\end{align}
The inequality in (\ref{eq:det-inequality}) follows from the following observation:
\begin{align}
\det\left[\mathbf{I}+\mu N\textsc{snr}\mathbf{\Lambda}_{A}\mathbf{\Lambda}_{K_{z}^{-1}}\right] &= \prod_i(1 + \mu N\textsc{snr} \lambda_{A,i} \lambda_{K_{z}^{-1},i})
\\
&\le \prod_i(1 + \mu N\textsc{snr} \lambda_{A,i}) \label{eq:det-inequality2}
\\
&= \det\left[\mathbf{I}+\mu N\textsc{snr}\mathbf{\Lambda}_{A}\right].
\end{align}
Above, $\lambda_A$ and $\lambda_{K_{z}^{-1}}$ denote the eigenvalues of $\mathbf{A}$ and $\K_z^{-1}$, respectively. The inequality in (\ref{eq:det-inequality2}) follows from the fact that the eigenvalues of $\K_z^{-1}$ are smaller than 1, i.e., $\frac{\sigma_{n}^{2}}{N(\sigma_{n}^{2}+\sigma_{s}^{2})}\le\lambda_{K_z^{-1},i}\le 1$ as mentioned before, and the fact that $\lambda_{A,i} \ge 0$ which is due to the positive semi-definiteness of $\mathbf{H}\mathbf{K}_{x_{1}}\mathbf{H}^{\dag}$ \footnote{The positive semi-definiteness can be easily seen from the following simple argument. For any vector $\mathbf{a}$, we can write $\mathbf{a}^{\dd}\mathbf{H}\mathbf{K}_{x_{1}}\mathbf{H}^{\dag} \mathbf{a} = \mathbf{b}^{\dd}\mathbf{K}_{x_{1}} \mathbf{b} \ge 0 $, where we have defined $\mathbf{b} = \mathbf{H}^{\dag} \mathbf{a}$ and used the fact that $\K_{x_1}$ is positive semi-definite.}.
From the inequality established through (\ref{eq:det-inequality0}) -- (\ref{eq:det-inequality1}),   we see that, in scenario 3, the transmission rate is less than the capacity (i.e., $r_{1} \leq C_{3}$). Hence, although reliable transmission is achieved at the rate of $r_1$, channel is not
fully utilized due to the false alarm in channel sensing. On the other hand, in a similar manner, it can be shown that in scenario 2, we
have the transmission rate $r_2$ exceeding the channel capacity $C_{2}$ because sensing
has not led to the successful detection of the active PUs, and the PUs' interference on the SUs'
signals is not taken into account. 
In this case, we assume that reliable communication cannot be achieved. Hence, the transmission
rate is effectively zero, and retransmission is required in scenario 2.
In the other three scenarios, communication is performed reliably. These four scenarios or
equivalently states are depicted in Figure $\ref{fig:fig1}$. Following the discussion above, we assume that the channel is ON in
states 1,3, and 4, in which data is sent reliably, and is OFF in state 2.

\begin{figure}
\begin{center}
\includegraphics[width =\figsize\textwidth]{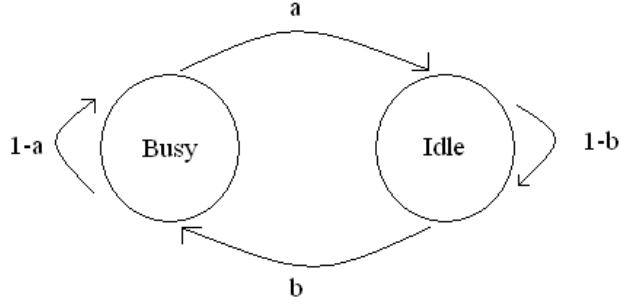}
\caption{Two-state Markov model for the PU activity.}
\label{Markov}
\end{center}
\end{figure}

Next, we determine the state-transition probabilities. We use $p_{ij}$ to denote the transition
probability from state $i$ to state $j$ as seen in Fig. $\ref{fig:fig1}$. Due to the block fading
assumption, state transitions occur every $T$ seconds. We also assume that PU activity does not change within each frame. We consider a two-state Markov model to describe the transition of the PU activity between the frames. This Markov model is depicted in Figure \ref{Markov}.
Busy state indicates that the channel is occupied by the PUs, and idle state indicates that there is no PU present in the channel. Probability of transitioning from busy state to idle state is denoted by $\textit{a}$, and the probability of transitioning from idle state to busy state is denoted by $\textit{b}$. Let us first consider in detail the probability of staying in the topmost ON state in Fig. $\ref{fig:fig1}$. This probability, denoted by $p_{11}$, is given by
\begin{align}
p_{11}&= \Pr\left\{ \substack{\text{channel is busy and is detected busy} \\ \text{in the $l^{th}$ frame}} \,\, \Big | \,\, \substack{\text{channel is busy and is detected busy} \\ \text{in the $(l-1)^{th}$ frame}} \right\} \label{pro11_1}
\\
&= \Pr\left\{ \substack{\text{channel is busy } \\ \text{in the $l^{th}$ frame}} \,\, \Big | \,\, \substack{\text{channel is busy } \\ \text{in the $(l-1)^{th}$ frame}} \right\} \times \Pr\left\{ \substack{\text{channel is detected busy} \\ \text{in the $l^{th}$ frame}} \,\, \Big | \,\, \substack{\text{channel is busy} \\ \text{in the $l^{th}$ frame}} \right\} \nonumber
\\
&=(1-a)P_{d}
\end{align}
where $P_{d}$ is the probability of detection in channel sensing. Channel being busy in the $l^{th}$ frame depends only on channel being busy in the $(l-1)^{th}$ frame and not on the other events in the condition. Moreover, since channel sensing is performed individually in each frame without any dependence on the channel sensing decision and PU activity in the previous frame, channel being detected as busy in the $l^{th}$ frame depends only on the event that the channel is actually busy in the $l^{th}$ frame.

Similarly, the probabilities for transitioning from any state to state 1 (topmost ON state) can be expressed as
\begin{align}\label{p11ler}
p_{b1}=p_{11}&=p_{21}=(1-a)P_{d}\quad\textrm{and}\quad p_{i1}=p_{31}=p_{41}=bP_{d}.
\end{align}
Note that we have common expressions for the transition probabilities in cases in which the originating state has a busy channel (i.e., states 1 and 2) and in cases in which the originating state has an idle channel (i.e., states 3 and 4).

In a similar manner, the remaining transition probabilities are given by the following:

For all $b \in \{1,2\}$ and $i \in \{3,4\}$,
\begin{align}\label{prob2}
\begin{array}{ll}
p_{b2}=(1-a)(1-P_{d}),\quad\textrm{and}\quad p_{i2}=b(1-P_{d}),\\
p_{b3}=aP_{f},\quad\textrm{and}\quad p_{i3}=(1-b)P_{f},\\
p_{b4}=a(1-P_{f}),\quad\textrm{and}\quad p_{i4}=(1-b)(1-P_{f}).
\end{array}
\end{align}

Now, we can easily see that the $4\times4$ state transition matrix can be expressed as
\begin{align}\label{R}
R=\left(
  \begin{array}{cccc}
    p_{11} & . & . & p_{14} \\
    p_{21} & . & . & p_{24} \\
    p_{31} & . & . & p_{34} \\
    p_{41} & . & . & p_{44} \\
  \end{array}
\right)=\left(
  \begin{array}{cccc}
    p_{b1} & . & . & p_{b4} \\
    p_{b1} & . & . & p_{b4} \\
    p_{i1} & . & . & p_{i4} \\
    p_{i1} & . & . & p_{i4} \\
  \end{array}
\right).
\end{align}

\subsection{Effective Capacity}
In \cite{Wu}, Wu and Negi defined the effective capacity as the maximum constant arrival rate that a given service process can support in order to guarantee a statistical QoS requirement specified by the QoS exponent $\theta$. If we define $Q$ as the stationary queue length, then $\theta$ is defined as the decay rate of the tail distribution of the queue length $Q$:
\begin{equation}\label{decayrate}
\lim_{q\rightarrow \infty}\frac{\log \Pr(Q\geq q)}{q}=-\theta.
\end{equation}
Hence, we have the following approximation for the buffer violation probability for large $q_{max}$: $\Pr(Q\geq q_{max})\approx e^{-\theta q_{max}}$. Therefore, larger $\theta$ corresponds to more strict QoS constraints, while the smaller $\theta$ implies looser constraints. In certain settings, constraints on the queue length can be linked to limitations on the delay and hence delay-QoS constraints. It is shown in \cite{Liu} that $\Pr\{D\geq d_{max}\}\leq c\sqrt{\Pr\{Q\geq q_{max}\}}$ for constant arrival rates, where $D$ denotes the steady-state delay experienced in the buffer. In the above formulation, $c$ is a positive constant, $q_{max}=ad_{max}$ and $a$ is the source arrival rate. Therefore, effective capacity provides the maximum arrival rate when the system is subject to statistical queue length or delay constraints  in the forms of $\Pr(Q \ge q_{\max}) \le e^{-\theta q_{max}}$ or $\Pr\{D \ge d_{\max}\} \le c \, e^{-\theta a \, d_{max}/2}$, respectively, for large thresholds $q_{\max}$ and $d_{\max}$. Since the average arrival rate is equal to the average departure rate when the queue is in steady-state \cite{ChangZajic}, effective capacity can also be seen as the maximum throughput in the presence of such constraints.

The effective capacity for a given QoS exponent $\theta$ is formulated as
\begin{equation}\label{exponent}
-\lim_{t\rightarrow \infty}\frac{1}{\theta t}\log_{e}\mathbb{E}\{e^{-\theta S(t)}\}=-\frac{\Lambda(-\theta)}{\theta}
\end{equation}
where $\Lambda(\theta) = \lim_{t\rightarrow \infty}\frac{1}{t}\log_{e}\mathbb{E}\{e^{\theta S(t)}\}$
is a function that depends on the logarithm of the moment generating function of $S(t)$,
$S(t)=\sum_{k=1}^{t}r(k)$ is the time-accumulated service process, and $\{r(k),k=1,2,\dots\}$ is
defined as the discrete-time, stationary and ergodic stochastic service process. Note that the
service rate in each transmission block is $r(k) = Tr_{1}$ if the cognitive system is in Scenario 1
or 3 at time $k$. Similarly, the service rate is $r(k) = Tr_{2}$ in Scenario 4. In the OFF state in
Scenario 2, the service rate is effectively zero.

Considering the effective rates in each scenario and the probabilities of the scenarios, we have
the following theorem.

\begin{theo} \label{theo:effective capacity}
For the CR channel with the aforementioned state transition model , the normalized
effective capacity in bits/s/Hz/dimension is given by
\begin{align}\label{effective capacity}
C_{E}(\textsc{snr},\theta)&=\max_{\substack{0 \le \mu \le 1\\0 \le P_2 \le \min\left\{P_{\max}, \frac{P_{int}}{P_d \mu + (1-P_d)}\right\}}}-\frac{1}{\theta
TBN}\log_{e}\mathbb{E}\bigg\{\frac{1}{2}\left[\left(p_{b1}+p_{i3}\right)e^{-\theta
Tr_{1}}+p_{i4}e^{-\theta Tr_{2}}+p_{b2}\right]\nonumber\\
&\frac{1}{2}\left\{\left[\left(p_{b1}-p_{i3}\right)e^{-\theta Tr_{1}}-p_{i4}e^{-\theta
Tr_{2}}+p_{b2}\right]^{2}+4\left(p_{i1}e^{-\theta Tr_{1}}+p_{i2}\right)\left(p_{b3}e^{-\theta
Tr_{1}}+p_{b4}e^{-\theta Tr_{2}}\right)\right\}^{1/2}\bigg\}
\end{align}
where $T$ is the frame duration over which the fading stays constant, $r_1$ and $r_2$ are the transmission rates given in (\ref{r1}) and (\ref{r2}), and $\{p_{bk},p_{il}\}$ for $k,l\in{1,2,3,4}$ are the state transition probabilities given in (\ref{p11ler}) and (\ref{prob2}).
\end{theo}

\emph{Proof:} See Appendix \ref{app:effective capacity}.

Note that above we have assumed that $\mathbf{H}$ is perfectly known at the
transmitter, which, equipped with this knowledge, can choose the input covariance matrices to maximize the instantaneous channel capacities as seen in (\ref{r1}) and (\ref{r2}). If, on the other hand, only statistical information related to $\mathbf{H}$ are known at
the transmitter, then the input covariance matrix can be chosen to maximize the effective capacity.
In that case, the normalized effective capacity will be expressed as
\begin{align}\label{capacity with H statistics}
C_{E}&(\textsc{snr},\theta)=\max_{\substack{0 \le \mu \le 1\\0 \le P_2 \le \min\left\{P_{\max}, \frac{P_{int}}{P_d \mu + (1-P_d)}\right\}}}
\max_{\substack{\mathbf{K}_{x_{1}},\mathbf{K}_{x_{2}}\succeq0\\\tr(\mathbf{K}_{x_{1}}) = \tr(\mathbf{K}
_{x_{2}}) = 1}}
-\frac{1}{\theta
TBN}\log_{e}\mathbb{E}\bigg\{\frac{1}{2}\left[\left(p_{b1}+p_{i3}\right)\Theta_{r_{1}}+p_{i4} \Theta_
{ r_ { 2 } } +p_ { b2
}\right]\nonumber\\
&+\frac{1}{2}\left\{\left[\left(p_{b1}-p_{i3}\right)\Theta_{r_{1}}-p_{i4}\Theta_{r_{2}}+p_{b2}\right]
^ {2}+4\left(p_{i1}\Theta_{r_{1}}+p_{i2}\right)\left(p_{b3}\Theta_{r_{1}}+p_{b4}\Theta_{r_{2}}
\right)\right\}^{1/2}\bigg\}\textrm{bits/s/Hz/dimension}
\end{align}
where $\Theta_{r_{1}}=e^{-\theta
TB\log_{2}\det\left[\mathbf{I}+\mu
N\textsc{snr}\mathbf{H}\mathbf{K}_{x_{1}}\mathbf{H}^{\dag}\mathbf{K}_{z}^{-1}\right]}$ and
$\Theta_{r_{2}}=e^{-\theta TB\log_{2}\det\left[\mathbf{I}+N\textsc{snr}\mathbf{H}\mathbf{K}_{x_{2}}\mathbf{H}^{\dag}\right]}$.
Now, the input covariance matrices are selected to maximize the effective rate.
For given $\mu$ and $P_2$, and for given input covariance matrices $\mathbf{K}_{x_{1}}$ and $\mathbf{K}_{x_{2}}$, we express the effective rate as
\begin{align}\label{effective rate}
R_{E}&(\textsc{snr},\theta)=-\frac{1}{\theta TBN}\log_{e}\mathbb{E}\bigg\{
\frac{1}{2}\left[\left(p_{b1}+p_{i3}\right)\Theta_{r_{1}}+p_{i4}\Theta_{r_{2}}+p_{b2
}\right]\nonumber\\
&+\frac{1}{2}\left\{\left[\left(p_{b1}-p_{i3}\right)\Theta_{r_{1}}-p_{i4}\Theta_{r_{2}}+p_{b2}\right]
^ {2}+4\left(p_{i1}\Theta_{r_{1}}+p_{i2}\right)\left(p_{b3}\Theta_{r_{1}}+p_{b4}\Theta_{r_{2}}
\right)\right\}^{1/2}\bigg\}\textrm{bits/s/Hz/dimension}.
\end{align}

\subsection{Ergodic Capacity}
As $\theta$ vanishes, the QoS constraints become loose and it can be easily verified that the effective capacity approaches the ergodic channel capacity, i.e.,
\begin{align}\label{capacity teta 0}
\hspace{-.1cm}\lim_{\theta\rightarrow0}C_{E}(\textsc{snr},\theta)&=\frac{1}{N}\max_{\substack{0 \le \mu \le 1\\0 \le P_2 \le \min\left\{P_{\max}, \frac{P_{int}}{P_d \mu + (1-P_d)}\right\}}}
\frac{bP_{d}+aP_{f}}{a+b}\mathbb{E}\left\{\max_{\substack{\mathbf{K}_{x_{1}}\succeq0\\\tr(\mathbf{K}_{x_{1}}) = 1}}\log_2\det\left[\mathbf{I}+\mu
N\textsc{snr}\mathbf{H}\mathbf{K}_{x_{1}}\mathbf{H}^{\dag}\mathbf{K}_{z}^{-1}\right]\right\}\nonumber\\
&\hspace{5.4cm}+\frac{a(1-P_{f})}{a+b}\mathbb{E}\left\{\max_{\substack{\mathbf{K}_{x_{2}}\succeq0\\\\tr(\mathbf{K}
_{x_{2}}) = 1}}\log_2\det\left[\mathbf{I}+
N\textsc{snr}\mathbf{H}\mathbf{K}_{x_{2}}\mathbf{H}^{\dag}\right]\right\}.
\end{align}
%
In order to gain further insight on the ergodic capacity expression, we note the following:
\begin{align}
\Pr\{\substack{\text{channel is}\\\text{detected busy}}\} &= \Pr\{\substack{\text{channel}\\\text{is busy}}\}\Pr\{\substack{\text{channel is }\\\text{detected busy}}\mid \substack{\text{channel}\\\text{is busy}}\} + \Pr\{\substack{\text{channel}\\\text{is idle}}\}\Pr\{\substack{\text{channel is }\\\text{detected busy}}\mid \substack{\text{channel}\\\text{is idle}}\}
\\
&= \frac{b}{a+b} P_d + \frac{a}{a+b} P_f
\\
&= \frac{bP_d + aP_f}{a+b} \label{eq:probofbusydet}
\end{align}
where we used the fact that for the two-state Markov model of the PU activity depicted in Fig. \ref{Markov}, the probability of being in the busy state is $a/(a+b)$. Similarly, we have
\begin{align}
\Pr\{\substack{\text{channel is idle and}\\\text{is detected idle}}\} = \Pr\{\substack{\text{channel}\\\text{is idle}}\}\Pr\{\substack{\text{channel is }\\\text{detected idle}}\mid \substack{\text{channel}\\\text{is idle}}\} = \frac{a}{a+b} (1-P_f). \label{eq:probofidleandidledet}
\end{align}
Recall that when the channel is detected busy, the transmitter sends the data at the rate $r_{1}$ given in (\ref{r1}), and the transmission is successful because we are in either state 1 or 3 (of the state transition model in Fig. \ref{fig:fig1}) which are both ON. If the channel is idle and is detected idle, then we are in state 4, which is also ON, and data is transmitted successfully at the rate $r_2$ given in (\ref{r2}). On the other hand, when the channel is busy but is detected idle, the rate $r_2$ cannot be supported by the channel and reliable communication cannot be achieved. Consequently, in this scenario (which is state 2 in Fig. \ref{fig:fig1}), the successful transmission rate is zero. From this discussion, we immediately realize that the ergodic capacity in (\ref{capacity teta 0}) is proportional to the average of these transmission rates weighted by the probabilities of the corresponding scenarios.

%


\section{Energy Efficiency in the Low-Power Regime}\label{effective capacity low power}

In this section, we investigate the performance of cognitive MIMO transmissions in the low-power regime. For this analysis, we consider the following second-order low-{\snr} expansion of the effective capacity:
\begin{equation}\label{effective capacity second order}
C_{E}(\textsc{snr},\theta)=\dot{C}_{E}(0,\theta)\textsc{snr}+\ddot{C}_{E}(0,\theta)\frac{\textsc{snr}^{2}}{2}+
o(\textsc{snr}^{2})
\end{equation}
where $\dot{C}_{E}(0,\theta)$ and $\ddot{C}_{E}(0,\theta)$ denote the first and second derivatives of the effective capacity with respect to $\textsc{snr}$ at $\textsc{snr}=0$. Note that the above expansion provides an accurate approximation of the effective capacity at low $\snr$ levels.

The benefits of a low-$\snr$ analysis are mainly twofold. First, operating at low power levels limits the interference inflicted on the PUs which is an important consideration in practice. Secondly, as will be seen below, energy efficiency improves as one lowers the transmission power. Hence, in this section, we consider a practically appealing and ambitious scenario in which cognitive users, in addition to their primary goal of efficiently utilizing the spectrum by filling in the spectrum holes, strive to operate energy efficiently while at the same time severely limiting the interference they cause on the PUs.

For the energy efficiency analysis, we adopt the \emph{energy per bit} given by
\begin{gather}
\frac{E_b}{N_0} = \frac{\textsc{snr}}{C_{E}(\textsc{snr},\theta)},
\end{gather}
as the performance metric. It is shown in \cite{Verdu} that the bit energy requirements diminish as {\snr} is lowered and the minimum energy per bit is achieved as $\snr$ vanishes, i.e.,
\begin{equation}\label{Bit Energy}
\frac{E_{b}}{N_{0}}_{\min}=\lim_{\textsc{snr}\rightarrow0}\frac{\textsc{snr}}{C_{E}(\textsc{snr}, \theta)}=\frac{1}{\dot{C}_{E}(0,\theta)}.
\end{equation}
Note that $\frac{E_{b}}{N_{0}}_{\min}$ is characterized only by the first derivative $\dot{C}_{E}(0,\theta)$. At $\frac{E_{b}}{N_{0}}_{\min}$, the slope $\mathcal{S}_{0}$ of the effective capacity versus $E_{b}/N_{0}$ (in dB) curve is defined as \cite{Verdu}
\begin{equation}\label{Slope}
\mathcal{S}_{0}=\lim_{\frac{E_{b}}{N_{0}}\downarrow\frac{E_{b}}{N_{0}}_{\min}}\frac{C_{E}\left(\frac{E_{b}}{N_{0}}\right)}{10\log_{10}\frac{E_{b}}{N_{0}} -10\log_{10}\frac{E_{b}}{N_{0}}_{\min}}10\log_{10}2.
\end{equation}
Considering the expression for the effective capacity, the wideband slope can be found from \cite{Verdu}
\begin{equation}\label{slope2}
\mathcal{S}_{0}=\frac{2\left[\dot{C}_{E}(0,\theta)\right]^{2}}{-\ddot{C}_{E}(0,\theta)}\log_{e}2\quad \textrm{bits/s/Hz/(3 dB)/receive antenna}.
\end{equation}
Hence, the wideband slope is obtained from both the first and second derivatives at $\snr = 0$. The wideband slope $\mathcal{S}_0$ together with the minimum energy per bit $\frac{E_{b}}{N_{0}}_{\min}$ provide a linear approximation of the effective capacity as a function of the energy per bit in the low-$\snr$ regime and enable us to gain insight on the energy efficiency of cognitive transmissions.

The next result identifies the first derivative of the effective capacity and the minimum bit energy.

\begin{theo} \label{theo:first-deriv}
In the cognitive MIMO channel considered in this paper, the first derivative of the effective capacity with respect to $\snr$ at $\snr = 0$ is
\begin{equation}\label{derivative first effective capacity}
\dot{C}_{E}(0,\theta)=\frac{1}{\log_{e}2}\bigg\{\frac{bP_{d}+aP_{f}}{a+b}\mathbb{E}\left[\lambda_{\max}(\mathbf{H}^{\dag}\mathbf{K}_{z}^{-1}\mathbf{H})\right] +\frac{a(1-P_{f})}{a+b}\mathbb{E}\left[\lambda_{\max}(\mathbf{H}^{\dag}\mathbf{H})\right]\bigg\}.
\end{equation}
Consequently, the minimum energy per bit is given by
\begin{equation}\label{min-bit-energy}
\frac{E_{b}}{N_{0}}_{\min}=\frac{\log_{e}2}{\frac{bP_{d}+aP_{f}}{a+b}\mathbb{E}\left[\lambda_{\max}(\mathbf{H}^{\dag}\mathbf{K}_{z}^{-1}\mathbf{H})\right] +\frac{a(1-P_{f})}{a+b}\mathbb{E}\left[\lambda_{\max}(\mathbf{H}^{\dag}\mathbf{H})\right]}.
\end{equation}
\end{theo}

\emph{Proof:} See Appendix \ref{app:first-deriv}.

\begin{rem}
As detailed in the proof of Theorem \ref{theo:first-deriv}, the first derivative of the
effective capacity at $\snr = 0$ and hence the
minimum energy per bit is achieved by transmitting
in the maximal-eigenvalue
eigenspaces
of $\mathbf{H}^{\dag}\mathbf{K}_{z}^{-1}\mathbf{H}$ and $\mathbf{H}^{\dag}\mathbf{H}$,
when the channel is sensed as busy and idle,
respectively. For instance, input covariance
matrices in the cases of busy- and idle-sensed
channels can be chosen, respectively, as
\begin{equation}
\mathbf{K}_{x_{1}}=\mathbf{u}_{1}\mathbf{u}_{1}^{\dag}
\quad \text{ and } \quad
\mathbf{K}_{x_{2}}=\mathbf{u}_{2}\mathbf{u}_{2}^{\dag}
\end{equation}
where $\mathbf{u}_{1}$ and $\mathbf{u}_{2}$ are the unit-norm eigenvectors associated with the maximum eigenvalues $\lambda_{\max}(\mathbf{H}^{\dag}\mathbf{K}_{z}^{-1}\mathbf{H})$ and $\lambda_{\max}(\mathbf{H}^{\dag}\mathbf{H})$, respectively. Hence, beamforming in the eigenvector directions corresponding to the maximum eigenvalues of $\mathbf{H}^{\dag}\mathbf{K}_{z}^{-1}\mathbf{H}$ and $\mathbf{H}^{\dag}\mathbf{H}$ is optimal in terms of energy efficiency. Note that when the channel is sensed as busy, the possible interference arising from the primary users' transmissions is taken into account by incorporating $\mathbf{K}_{z}^{-1}$ into the transmission strategy. Note further that as shown in (\ref{eq:probofbusydet}) and (\ref{eq:probofidleandidledet}), $\frac{bP_{d}+aP_{f}}{a+b}$ is the probability of detecting the channel as busy, and $\frac{a(1-P_{f})}{a+b}$ is the probability that channel is idle and is detected as idle.
\end{rem}

\begin{rem}
The expressions in (\ref{derivative first effective capacity}) and (\ref{min-bit-energy}) do not depend on the QoS exponent $\theta$, indicating that the performance  in the low power regime as $\snr \to 0$ does not get affected by the presence of QoS requirements. Indeed, $\frac{E_{b}}{N_{0}}_{\min}$ in (\ref{min-bit-energy}) is the minimum energy per bit attained when no QoS constraints are imposed.
\end{rem}

\begin{rem}
It is also interesting to note that the sensing performance has an impact on the energy efficiency. In particular, we can immediately notice that $\frac{E_{b}}{N_{0}}_{\min}$ decreases with increasing detection probability $P_d$. Similarly, $\frac{E_{b}}{N_{0}}_{\min}$ decreases as the false alarm probability $P_f$ decreases. This can be seen by noticing that decreasing $P_f$ leads to an increased weight on $\mathbb{E}\left[\lambda_{\max}(\mathbf{H}^{\dag}\mathbf{H})\right]$ and a decreased weight on $\mathbb{E}\left[\lambda_{\max}(\mathbf{H}^{\dag}\mathbf{K}_{z}^{-1}\mathbf{H})\right] $, and noting that using Ostrowski's Theorem \cite[Theorem 4.5.9 and Corollary 4.5.11]{Matrix Analysis} and its extension to non-square transforming matrices in \cite[Theorems 3.2 and 3.4]{Higham}, we have
\begin{align}
\lambda_{\max}(\mathbf{H}^{\dag}\mathbf{K}_{z}^{-1}\mathbf{H}) \le \lambda_{\max}(\mathbf{K}_{z}^{-1}) \lambda_{\max}(\mathbf{H}^{\dag}\mathbf{H}) \le \lambda_{\max}(\mathbf{H}^{\dag}\mathbf{H})
\end{align}
where the last inequality follows from the property that $\lambda_{\max}(\mathbf{K}_{z}^{-1}) \le 1$.
\end{rem}

Since minimum energy per bit is a metric in the asymptotic regime in which $\snr$ vanishes, we next consider the wideband slope in order to identify the performance at low but nonzero $\snr$ levels. Wideband slope in (\ref{slope2}) depends on the both the first and second derivatives of the effective capacity at $\snr = 0$. In obtaining the second derivative, we essentially make use of the fact that the optimal input covariance matrices in the low $\snr$ regime, which are
required to achieve the minimum bit energy and hence the wideband slope, can be
expressed as
\begin{equation}
\mathbf{K}_{x_{1}}=\sum_{i=1}^{m_1}\kappa_{1i}\mathbf{u}_{1,i}\mathbf{u}_{1,i}^{\dag}
\quad \text{ and } \quad
\mathbf{K}_{x_{2}}=\sum_{i=1}^{m_2}\kappa_{2i}\mathbf{u}_{2,i}\mathbf{u}_{2,i}^{\dag}
\end{equation}
where $\kappa_{1i},\kappa_{2i}\in[0,1]$ are the weights satisfying $\sum_{i=1}^{m_1}\kappa_{1i}=1$ and $\sum_{i=1}^{m_2}\kappa_{2i}=1$, and $m_1 \ge 1$ and $m_2 \ge 1$ are the multiplicities of $\lambda_{\max}(\mathbf{H}^{\dag}\mathbf{K}_{z}^{-1}\mathbf{H})$ and $\lambda_{\max}(\mathbf{H}^{\dag}\mathbf{H})$, respectively. Moreover, $\{\mathbf{u}_{1,i}\}$ and $\{\mathbf{u}_{2,i}\}$ are the orthonormal eigenvectors that span the maximal-eigenvalue eigenspaces of $\mathbf{H}^{\dag}\mathbf{K}_{z}^{-1}\mathbf{H}$ and $\mathbf{H}^{\dag}\mathbf{H}$, respectively. Despite this characterization, obtaining a general closed-form expression for the second-derivative seems intractable and we concentrate on the special case in which $a + b = 1$. Note that this case represents a scenario where there is no memory in the two-state Markov model for the PU activity. Hence, for instance, transitioning from busy state to busy state has the same probability as transitioning from idle state to busy state.

\begin{theo} \label{theo:second-deriv}
In the special case in which the transition probabilities satisfy $a + b = 1$ in the two-state model for the PU activity, the second derivative of the effective capacity with respect to $\snr$ at $\snr = 0$ is
\begin{align}
\ddot{C}_{E}(0,\theta) &= \frac{\theta TBN}{\log_{e}^{2}2}\mathbb{E}^{2}\left[\ell_{1}\lambda_{\max}(\mathbf{H}^{\dag}\mathbf{K}_{z}^{-1}\mathbf{H})+\ell_{2} \lambda_{\max}(\mathbf{H}^{\dag}\mathbf{H})\right]\nonumber\\&-\frac{\theta TBN}{\log_{e}^{2}2}\mathbb{E}\left[\ell_{1}\lambda_{\max}^{2}(\mathbf{H}^{\dag}\mathbf{K}_{z}^{-1}\mathbf{H}) +\ell_{2}\lambda_{\max}^{2}(\mathbf{H}^{\dag}\mathbf{H})\right]\nonumber\\
&-\frac{N}{\log_{e}2}\mathbb{E}\left[\frac{\ell_{1}\lambda_{\max}^{2}(\mathbf{H}^{\dag}\mathbf{K}_{z}^{-1}\mathbf{H})}{m_{1}} +\frac{\ell_{2}\lambda_{\max}^{2}(\mathbf{H}^{\dag}\mathbf{H})}{m_{2}}\right]\nonumber
\end{align}
where $m_1$ and $m_2$ are the multiplicities of the eigenvalues $\lambda_{\max}(\mathbf{H}^{\dag}\mathbf{K}_{z}^{-1}\mathbf{H})$ and $\lambda_{\max}(\mathbf{H}^{\dag}\mathbf{H})$, respectively, and we have defined $\ell_{1}=(bP_{d}+aP_{f})$ and $\ell_{2}=a(1-P_{f})$.
The wideband slope is
\begin{align}\label{widebandslope}
\mathcal{S}_{0}=\frac{2\mathbb{E}^{2}\left[\ell_{1}\lambda_{\max,1}+\ell_{2}\lambda_{\max,2}\right]}{\theta TBN\left\{\mathbb{E}\left[\ell_{1}\lambda_{\max,1}^{2}+\ell_{2}\lambda_{\max,2}^{2}\right]-\mathbb{E}^{2} \left[\ell_{1}\lambda_{\max,1}+\ell_{2}\lambda_{\max,2}\right]\right\}+N\mathbb{E}\left[\frac{\ell_{1}\lambda_{\max,1}^{2}}{m_{1}} +\frac{\ell_{2}\lambda_{\max,2}^{2}}{m_{2}}\right]\log_{e}2}.
\end{align}
where we used the notation $\lambda_{\max,1} = \lambda_{\max}(\mathbf{H}^{\dag}\mathbf{K}_{z}^{-1}\mathbf{H})$ and $\lambda_{\max,2} = \lambda_{\max}(\mathbf{H}^{\dag}\mathbf{H})$.

\end{theo}

\emph{Proof:} See Appendix \ref{app:second-deriv}.

%
%

Unlike the minimum energy per bit, second derivative and the wideband slope depend on the QoS exponent $\theta$. In particular, we immediately notice that as $\theta$ increases (i.e., the QoS constraints become more stringent), wideband slope decreases, worsening the energy efficiency. Note that lower slopes imply that the same throughput is attained at an increased level of energy per bit.

When we have equal power allocation, i.e., $\mathbf{K}_{x}=\frac{1}{M}\mathbf{I}$, and with the assumption that $\mathbf{s}$ with dimension $N\times1$ is a zero-mean Gaussian random vector with a covariance matrix $\mathbb{E}\{\mathbf{s}\mathbf{s}^{\dag}\}=\sigma_{s}^{2}\mathbf{I}$ where $\mathbf{I}$ is the identity matrix, we can obtain
\begin{equation}\label{obtained 1_1}
\frac{E_{b}}{N_{0}}_{min}=\frac{\log_{e}2}{\left(\frac{\ell_{1}}{\sigma_{s}^{2}}+\ell_{2}\right) \mathbb{E}\left[\tr(\mathbf{H}^{\dag}\mathbf{H})\right]}
\end{equation}
\begin{align}\label{obtained 2_1}
\mathcal{S}_{0}=\frac{2\left(\frac{\ell_{1}}{\sigma_{s}^{2}}+\ell_{2}\right)^{2}\mathbb{E}^{2}\left[\tr(\mathbf{H}^{\dag}\mathbf{H})\right]}{\theta TBN\left\{\left[\frac{\ell_{1}}{\sigma_{s}^{4}}+\ell_{2}\right]\mathbb{E}\left[\tr^{2}(\mathbf{H}^{\dag}\mathbf{H})\right] -\left[\frac{\ell_{1}}{\sigma_{s}^{2}}+\ell_{2}\right]^{2}\mathbb{E}^{2}\left[\tr(\mathbf{H}^{\dag}\mathbf{H})\right]\right\} +N\left[\frac{\ell_{1}}{\sigma_{s}^{4}}+\ell_{2}\right]\mathbb{E}\left[\tr\left((\mathbf{H}^{\dag}\mathbf{H})^{2}\right)\right]\log_{e}2}.
\end{align}
Now, assuming that $\mathbf{H}$ has independent zero-mean unit-variance complex Gaussian random entries, we have \cite{Lozano}
\begin{align}\label{HHHH}
\mathbb{E}\left[\tr(\mathbf{H}^{\dag}\mathbf{H})\right]=NM,\quad \mathbb{E}\left[\tr^{2}(\mathbf{H}^{\dag}\mathbf{H})\right]=NM(NM+1),\quad \mathbb{E}\left[\tr\left((\mathbf{H}^{\dag}\mathbf{H})^{2}\right)\right]=NM(N+M).
\end{align}
Using these facts, we can write the following minimum bit energy and wideband slope expressions for the case of uniform power allocation:
\begin{equation}\label{New_Bit_Energy}
\frac{E_{b}}{N_{0_{min}}}=\frac{\log_{e}2}{\left(\frac{\ell_{1}}{\sigma_{s}^{2}}+\ell_{2}\right)NM}
\end{equation}
\begin{equation}\label{New_Wideband_Slope}
\mathcal{S}_{0}=\frac{2\left(\frac{\ell_{1}}{\sigma_{s}^{2}}+\ell_{2}\right)^{2}M^{2}}{\theta TB\left\{\left[\frac{\ell_{1}}{\sigma_{s}^{4}}+\ell_{2}\right]M(NM+1)- \left[\frac{\ell_{1}}{\sigma_{s}^{2}}+\ell_{2}\right]^{2}M^{2}\right\} +\left[\frac{\ell_{1}}{\sigma_{s}^{4}}+\ell_{2}\right]M(N+M)\log_{e}2}.
\end{equation}

\section{Numerical Results}\label{Numeric}
In our numerical results, we consider a Rayleigh fading channel model where the components of the channel matrix $\mathbf{H}$ are independent and identically distributed (i.i.d.) zero-mean, unit variance, circularly symmetric Gaussian random variables. Moreover, we assume that input covariance matrix is $\mathbf{K}_{x}=\frac{1}{M}\mathbf{I}$ and that the components of the received signal coming from PUs are i.i.d. and have a variance $\sigma_{s}^{2}$ so that $\mathbf{K}_{z}=\frac{\sigma_{s}^{2}+\sigma_{n}^{2}}{\sigma_{s}^{2}}\mathbf{I}$.

Furthermore, as the objective function we consider the effective rate which is given as
\begin{align}\label{effective rate numerical}
R_{E}(\textsc{snr},\theta)=-\frac{1}{\theta TB}\log_{e}\mathbb{E}\bigg\{& \ell_{1}e^{-\theta TB\log_{2}\det\left[ \mathbf{I}+\frac{\mu N\sigma_{n}^{2}}{M(\sigma_{s}^{2} +\sigma_{n}^{2})}\textsc{snr}\mathbf{H}\mathbf{H}^{\dag}\right]} +\ell_{2}e^{-\theta TB\log_{2}\det\left[ \mathbf{I}+\frac{ N}{M}\textsc{snr}\mathbf{H}\mathbf{H}^{\dag} \right]}\nonumber\\&+b(1-P_{d})\bigg\}\textrm{bits/Hz/s}.
\end{align}
With these assumptions, effective rate can be computed by using the expression for the moment generating function of instantaneous mutual information given by Wang and Giannakis in \cite[Theorem 1]{Giannakis}. After adopting this expression into our effective rate formulation ($\ref{effective rate numerical}$), we obtain
\begin{align}\label{effective rate numerical extra}
R_{E}(\textsc{snr},\theta)=-\frac{1}{\theta TB}\log_{e}\bigg\{ &[bP_{d}+aP_{f}]\frac{\det\left[\mathbf{G}\left(\theta,\frac{\mu\sigma_{n}^{2}\textsc{snr}}{\sigma_{s}^{2}+\sigma_{n}^{2}}\right)\right]}{\prod_{i=1}^{k}\Gamma(d+i)\Gamma(i)} +a(1-P_{f})\frac{\det\left[\mathbf{G}\left(\theta,\textsc{snr}\right)\right]}{\prod_{i=1}^{k}\Gamma(d+i)\Gamma(i)}\nonumber\\&+b(1-P_{d})\bigg\}\textrm{bits/Hz/s}
\end{align}
where $k=\min(M,N)$, $d=\max(M,N)-\min(M,N)$, and $\Gamma(.)$ is the Gamma function. Here, $\mathbf{G}(\theta,\textsc{snr})$ is a $k\times k$ Hankel matrix whose $(m,n)^{th}$ component is
\begin{equation}\label{HAnkel}
g_{m,n}=\int_{0}^{\infty}\left(1+\frac{N}{M}\textsc{snr}\textit{z}\right)^{-\theta TB\log_{2}e}\textit{z}^{m+n+d-2}e^{-\textit{z}}d\textit{z} \qquad \text{for }{m,n=1,2,...,k}.
\end{equation}

%

\begin{figure}
\begin{center}
\includegraphics[width =\figsize\textwidth]{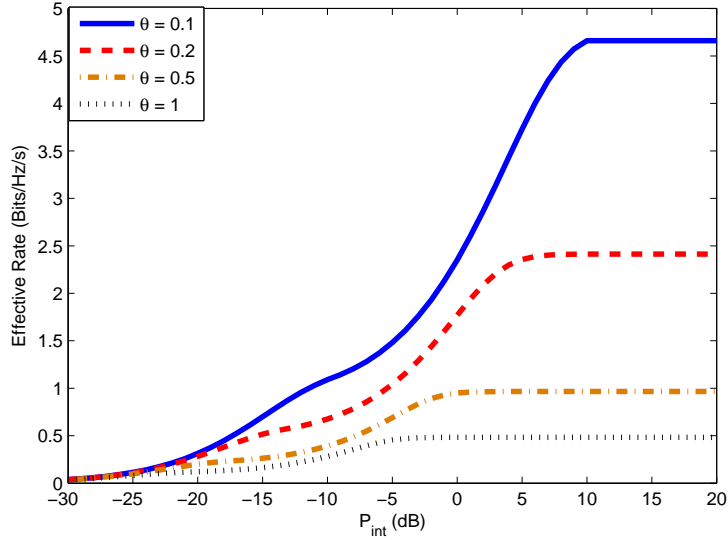}
\caption{Effective rate vs. power interference threshold, $P_{int}$ for different values of QoS exponent, $\theta$. $M = N = 3$.}
\label{fig:fig4-5}
\end{center}
\end{figure}

\begin{figure}
\begin{center}
\includegraphics[width =\figsize\textwidth]{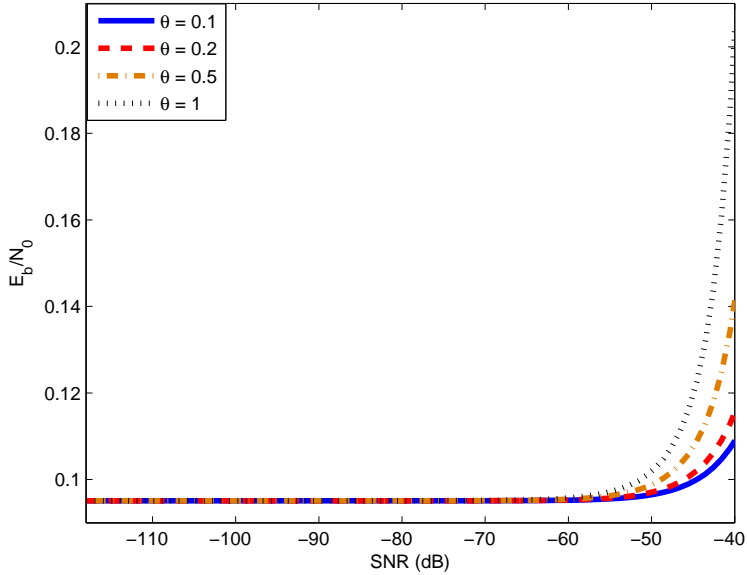}
\caption{Energy-per-bit, $\frac{E_{b}}{N_{0}}$, vs. SNR (dB) for different QoS exponent, $\theta$, values, respectively. $M = 3$ and $N = 3$.}
\label{fig:fig7}
\end{center}
\end{figure}

\begin{figure}
\begin{center}
\includegraphics[width =\figsize\textwidth]{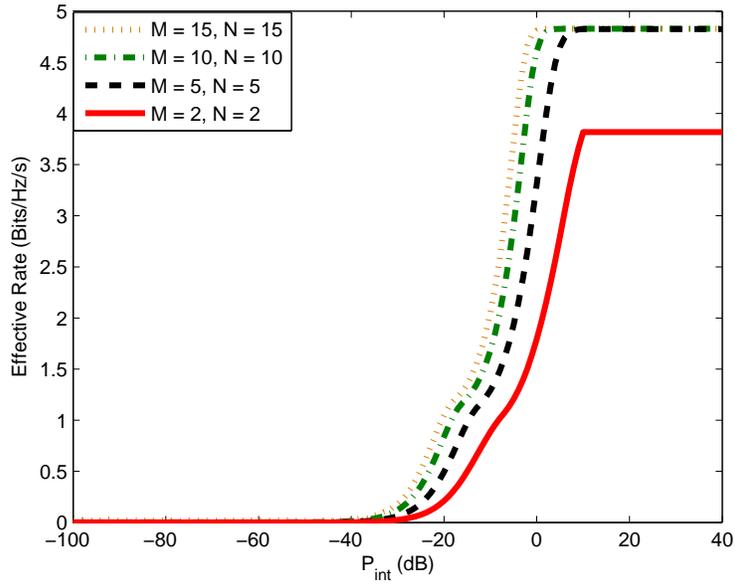}
\caption{Effective rate vs. power interference threshold, $P_{int}$ for different numbers of transmit and receive antennas, $M$ and $N$, respectively. $\theta = 0.1$.}
\label{fig:fig6}
\end{center}
\end{figure}

In our numerical results, we assume $T=0.1$ s, $B=100$ Hz, $\sigma_{n}^{2}=\sigma_{s}^{2}=1$, $P_{d}=0.92$, $P_{f}=0.21$, and $P_{\max}=10$ dB. In Figure \ref{fig:fig4-5}, we plot the effective rate as a function of $P_{int}$ for different values of the QoS exponent, $\theta$. In this figure, number of transmit and number of receive antennas are both 3, i.e., $ M = N = 3$. When the interference power threshold is low, the optimal ratio of power level $P_{1}$ to the power level $P_{2}$ is very small, i.e., $\mu = P_1/P_2 \sim 0$. Therefore, there is almost no transmission when the channel is detected as busy. Note in this case that false alarms lead to almost no transmission even if the channel is not occupied by the PUs. In addition, from (\ref{eq:P2upperbound}), we see that if the detection probability $P_d < 1$, then $P_2$, the transmission power when the channel is sensed as idle, scales with $P_{int}$ if $P_{int}$ is sufficiently small. Consequently, we see in Fig. \ref{fig:fig4-5}, the throughput diminishes to zero as $P_{int}$ gets smaller. On the other hand, as $P_{int}$ increases beyond a certain threshold, we observe that the effective rate becomes fixed due to OFF state (the state in which there is no data transmission and/or unreliable transmission), which becomes dominant in the effective rate expression, and the fact that even if $P_{int}$ is very high or there is no interference power threshold, average peak power, $P_{\max}$, limits the transmission powers. Another remark regarding the plots in Fig. \ref{fig:fig4-5} is that, as expected, the higher the QoS exponent $\theta$ (or equivalently the more strict the QoS constraints), the smaller the effective rate is. In Fig. \ref{fig:fig7}, we plot the corresponding energy-per-bit requirements, $\frac{E_{b}}{N_{0}}$,  as a function of SNR. Confirming our results, we observe that the minimum bit energy given in (\ref{New_Bit_Energy}) is indeed approached as SNR is diminished, and since the minimum energy per bit is independent of $\theta$, all curves converge as SNR vanishes. In Figure \ref{fig:fig6}, we plot the effective rate for different numbers of transmit and receive antennas as a function of $P_{int}$. We  set $\theta=0.1$. We observe that increasing the number of antennas beyond a certain level does not improve the transmission quality for higher values of $P_{int}$. On the other hand, for smaller values of $P_{int}$ in the range [-30dB, 0dB] (i.e., under relatively stringent interference constraints), with higher number of antennas, improvements in the throughput can be realized.

\begin{figure}
\begin{center}
\includegraphics[width =\figsize\textwidth]{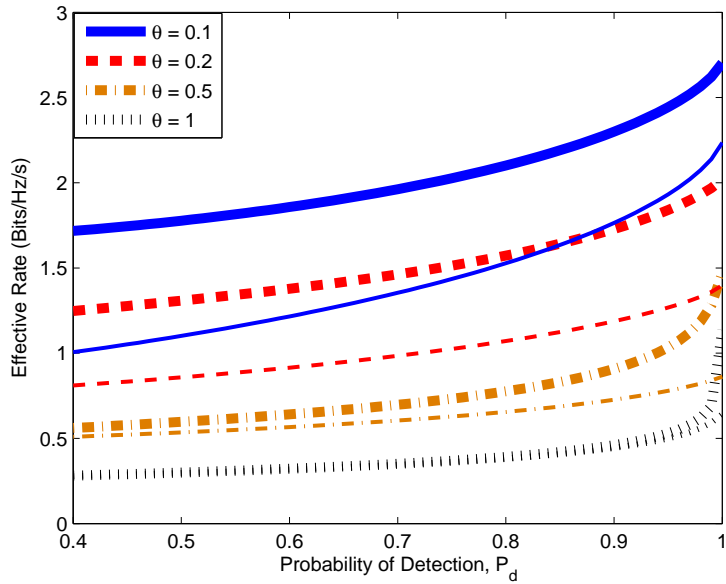}
\caption{Effective rate vs. probability of detection, $P_{d}$ for different QoS exponent, $\theta$, values, respectively. $M = 3$, $N = 3$, and $P_{int} = 0$ dB. For the thick lines, we have $P_{f}=0.21$ and for thin lines, we have $P_{f}=0.1$.}
\label{fig:fig8}
\end{center}
\end{figure}

\begin{figure}
\begin{center}
\includegraphics[width =\figsize\textwidth]{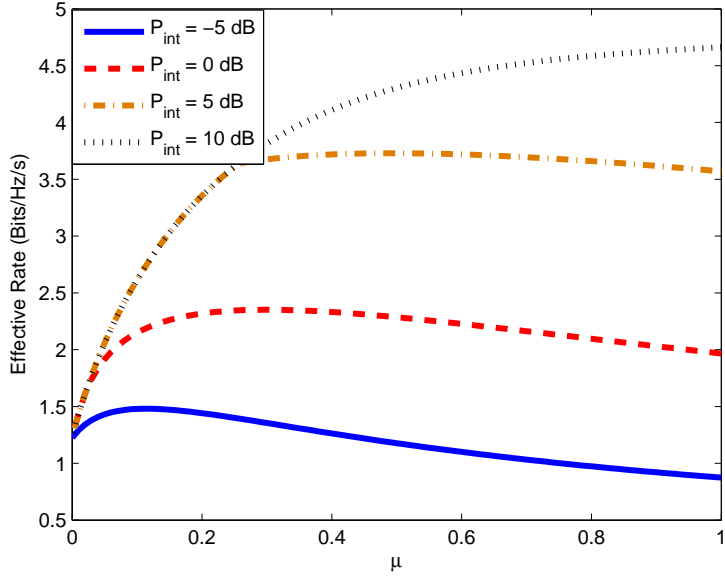}
\caption{Effective rate vs. power ratio, $\mu=\frac{P_{1}}{P_{2}}$, for different power interference values, $P_{int}$. $\theta = 0.1$, $M=3$ and $N=3$.}
\label{fig:fig9}
\end{center}
\end{figure}

\begin{figure}
\begin{center}
\includegraphics[width =\figsize\textwidth]{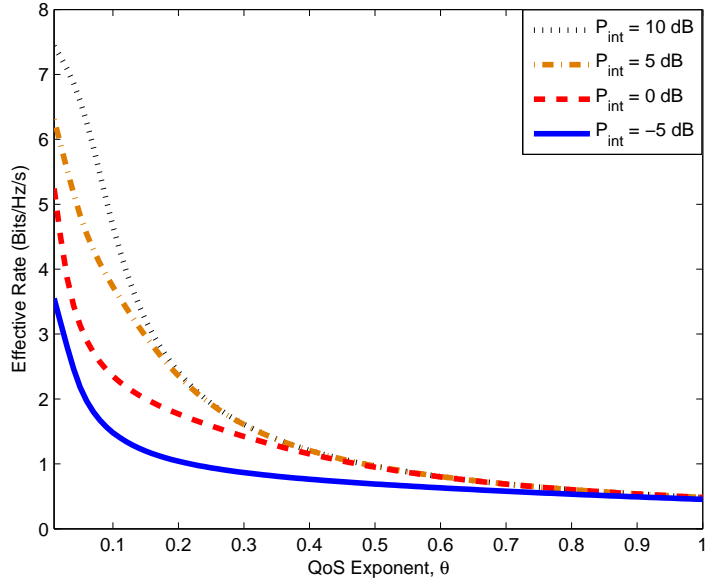}
\caption{Effective rate vs. QoS exponent, $\theta$, for different power interference values, $P_{int}$ at optimal $\mu$ values. $M=3$ and $N=3$.}
\label{fig:fig10}
\end{center}
\end{figure}

In Fig. \ref{fig:fig8}, we plot the effective rate as a function of probability of detection, $P_d$, when $P_{int} = 0$ dB. In this figure, we observe the impact of channel sensing performance on the throughput of cognitive MIMO transmissions. The curves with thick lines are obtained when probability of false alarm is $P_{f}=0.21$. Curves with thin lines are obtained when $P_f = 0.1$. With the increasing $P_{d}$, the effective rate increases as a result of efficient power allocation when the channel is sensed as idle. The interference caused by the primary users is controlled by allocating less power when the channel is sensed as busy. However, since the optimal power ratio $\mu = P_{1} / P_{2}$ depends on the value of the detection probability, $P_{d}$, the power allocated to transmission when the channel is sensed as idle decreases with the increasing $P_{d}$ but does not go to zero, which is because of the non-zero probability of false alarm, $P_{f}$. Therefore, we also observe that with the decreasing probability of false alarm, the effective rate decreases due to less power allocated when the channel is sensed as busy. Furthermore, in Fig. \ref{fig:fig9}, we plot the effective rate as a function of power ratio $\mu$ for different power interference values, $P_{int}$. We observe that with decreasing $P_{int}$, the optimal $\mu$ is decreasing for the aforementioned $P_{d}$ and $P_{f}$ values. Note that the optimal $\mu$ is 1 when $P_{int}=P_{max}$. Finally, in Figure \ref{fig:fig10}, we plot the effective rate as a function of the QoS exponent, $\theta$. As expected with the increasing $\theta$ values, the effective rate is decreasing due to more strict buffer/delay constraints. We also note that smaller $P_{int} $ and hence more strict interference constraints lead to reduced throughput for smaller values of $\theta$. On the other hand, if $\theta$ is large, the impact of $P_{int}$ lessens and curves converge.

\section{Conclusion}\label{Conclusion}
In this paper, we have investigated the throughput and energy efficiency of cognitive MIMO wireless communication systems operating under queuing constraints, interference limitations, and imperfect channel sensing. We have considered effective capacity and rate as our throughput metrics and formulated them in terms of instantaneous transmission rates and state transition probabilities, which in turn depend on the primary user activities and sensing reliability. Through numerical results, we have investigated the impact of QoS and interference constraints and sensing performance, and the benefit of multiple antenna transmissions. For the energy efficiency analysis, we have studied the effective capacity in the low-power regime. We have obtained expressions for the first and second derivatives of the effective capacity. We have determined the minimum energy per bit required in the cognitive MIMO system. We have remarked that while the minimum energy per bit does not get affected by the presence of the QoS constraints, it decreases as the channel sensing reliability improves. We have seen that the second derivative and the wideband slope depend on the QoS exponent $\theta$. We have also shown that the minimum energy per bit and wideband slope are achieved by performing beamforming in the maximal-eigenvalue eigenspace of the matrices $\mathbf{H}^{\dag}\mathbf{K}_{z}^{-1}\mathbf{H}$ and $\mathbf{H}^{\dag}\mathbf{H}$.


\appendix

\subsection{Proof of Theorem \ref{theo:effective capacity}} \label{app:effective capacity}

The proof follows along similar lines as in \cite{paper3} in which a single-antenna case with channel uncertainty is studied. In \cite[Chap. 7, Example 7.2.7]{Performance}, it is shown for Markov modulated processes that
\begin{gather} \label{eq:theta-envelope}
\frac{\Lambda(\theta)}{\theta} = \frac{1}{\theta} \log_e sp(\phi(\theta)R)
\end{gather}
where $sp(\phi(\theta)R)$ is the spectral radius (i.e., the maximum of the absolute values of the
eigenvalues) of the matrix $\phi(\theta)R$, $R$ is the transition matrix of the underlying Markov
process, and $\phi(\theta) = \text{diag}(\phi_1(\theta), \ldots, \phi_F(\theta))$ is a diagonal
matrix whose components are the moment generating functions of the processes in $F$ states. The rates
supported by the CR channel with the state transition model described above can be seen
as a Markov modulated process and hence the setup considered in \cite{Performance} can be immediately
applied to our setting. Note that the transmission rates are random in each state in the cognitive
channel. Therefore, the corresponding moment generating functions are
$\phi_{1}(\theta)=\phi_{3}(\theta)=\mathbb{E}\{e^{T\theta r_{1}}\}$,
$\phi_{4}(\theta)=\mathbb{E}\{e^{T\theta r_{2}}\}$ and $\phi_{2}(\theta)=1$. Then, using \eqref{R},
we can write
\begin{align}
\phi(\theta)R=\left[
\begin{array}{cccc}
\phi_{1}(\theta)p_{b1} & . & . & \phi_{1}(\theta)p_{b4} \\
\phi_{2}(\theta)p_{b1} &   &   & \phi_{2}(\theta)p_{b4} \\
\phi_{3}(\theta)p_{i1} &   &   & \phi_{3}(\theta)p_{i4} \\
\phi_{4}(\theta)p_{i1} & . & . & \phi_{4}(\theta)p_{i4} \\
\end{array}
\right]
=
\left[\begin{array}{cccc}
\mathbb{E}\{e^{T\theta r_{1}}\}p_{b1} & . & . & \mathbb{E}\{e^{T\theta r_{1}}\}p_{b4}\\
p_{b1} & . & . & p_{b4} \\
\mathbb{E}\{e^{T\theta r_{1}}\}p_{i1} &   &   & \mathbb{E}\{e^{T\theta r_{1}}\}p_{i4} \\
\mathbb{E}\{e^{T\theta r_{2}}\}p_{i1} &   &   & \mathbb{E}\{e^{T\theta r_{2}}\}p_{i4} \\
\end{array}
\right]
\end{align}
Since $\phi(\theta)R$ is a matrix with rank 2, we can readily find that
\begin{align}\label{eq:sp}
&sp(\phi(\theta)R) = \text{trace}(\phi(\theta)R)\nonumber\\
&=\frac{1}{2}\left\{\phi_{1}(\theta)p_{b1}+\phi_{2}(\theta)p_{b2}
+\phi_{3}(\theta)p_{i3}+\phi_{4}(\theta)p_{i4}\right\}\nonumber\\
&+\frac{1}{2}\left\{\left[\phi_{1}(\theta)p_{b1}+\phi_{2}(\theta)p_{b2}
-\phi_{3}(\theta)p_{i3}-\phi_{4}(\theta)p_{i4}\right]^{2}+4\left(\phi_{1}(\theta)p_{i1}+\phi_{2}
(\theta)p_{i2}\right)\left(\phi_{3}(\theta)p_{b3}+\phi_{4}(\theta)p_{b4}\right)\right\}^{1/2}
\nonumber
\\
&=\frac{1}{2}\left\{\left(p_{b1}+p_{i3}\right)\mathbb{E}\{e^{T\theta
r_{1}}\}+p_{i4}\mathbb{E}\{e^{T\theta
r_{2}}\}+p_{b2}\right\}\nonumber\\
&+\frac{1}{2}\left\{\left[\left(p_{b1}-p_{i3}\right)\mathbb{E}\{e^{T\theta
r_{1}}\}-p_{i4}\mathbb{E}\{e^{T\theta
r_{2}}\}+p_{b2}\right]^{2}+4\left(p_{i1}\mathbb{E}\{e^{T\theta
r_{1}}\}+p_{i2}\right)\left(p_{b3}\mathbb{E}\{e^{T\theta
r_{1}}\}+p_{b4}\mathbb{E}\{e^{T\theta
r_{2}}\}\right)\right\}^{1/2}.
\end{align}
Then, combining (\ref{eq:sp}) with (\ref{eq:theta-envelope}) and (\ref{exponent}), we obtain the
expression inside the maximization on the right-hand side of (\ref{effective capacity}). \hfill
$\square$

\subsection{Proof of Theorem \ref{theo:first-deriv}} \label{app:first-deriv}

We define a new function
\begin{align}\label{f}
f(\textsc{snr},\theta)&=\frac{1}{2}\left[\left(p_{b1}+p_{i3}\right)e^{-\theta Tr_{1}}+p_{i4}e^{-\theta Tr_{2}}+p_{b2
}\right]\nonumber\\
&+\frac{1}{2}\underbrace{\left\{\left[\left(p_{b1}-p_{i3}\right)e^{-\theta Tr_{1}}-p_{i4}e^{-\theta Tr_{2}}+p_{b2}\right]
^ {2}+4\left(p_{i1}e^{-\theta Tr_{1}}+p_{i2}\right)\left(p_{b3}e^{-\theta Tr_{1}}+p_{b4}e^{-\theta Tr_{2}}
\right)\right\}^{1/2}}_{\chi},
\end{align}
and we can write the effective rate in (\ref{effective rate}) as
\begin{equation}\label{effective rate_f}
R_{E}(\textsc{snr},\theta)=D\log_{e}\mathbb{E}\left[f(\textsc{snr},\theta)\right]
\end{equation}
where $D=-\frac{1}{\theta TBN}$.
The derivative of the effective rate with respect to $\snr$ will be
\begin{align}\label{derivative first 1}
&\dot{R}_{E}(\textsc{snr},\theta)=\frac{D}{\mathbb{E}\left[f(\textsc{snr},\theta)\right]}\mathbb{E}\left[\dot{f}(\textsc{snr},\theta)\right]
\end{align}
where
\begin{equation}\label{f first derivative}
\dot{f}(\textsc{snr},\theta)=-\theta T\alpha(\textsc{snr},\theta)\dot{r}_{1}e^{-\theta Tr_{1}}-\theta T\beta(\textsc{snr},\theta)\dot{r}_{2}e^{-\theta Tr_{2}}
\end{equation}
and $\alpha(\textsc{snr},\theta)=\frac{1}{2}(p_{b1}+p_{i3})+\frac{(p_{b1}-p_{i3})\left[(p_{b1}-p_{i3})e^{-\theta Tr_{1}}-
p_{i4}e^{-\theta Tr_{2}}+p_{b2}\right]}{2\chi}+\frac{p_{i1}\left(p_{b3}e^{-\theta Tr_{1}}+p_{b4}e^{-\theta Tr_{2}}\right) +
p_{b3}\left(p_{i1}e^{-\theta Tr_{1}}+p_{i2}\right)}{\chi}$, $\beta(\textsc{snr},\theta)=\frac{1}{2}p_{i4}-\frac{p_{i4}\left[\left(p_{b1}-p_{i3}\right)e^{-\theta T_{r_{1}}}-p_{i4}e^{-\theta Tr_{2}}+p_{b2}\right]}{2\chi}+\frac{p_{b4}\left(p_{i1}e^{-\theta Tr_{1}}+p_{i2}\right)}{\chi}$, and $\chi$ is defined in (\ref{f}).
Note that we can write $r_{1}$ and $r_{2}$ as
\begin{equation}\label{r1u_new}
r_{1}=\frac{B}{\log_{e}2}\sum_{i}\log_{e}\left[1+\mu N\textsc{snr}\lambda_{i}(\Phi_{1})\right]
\end{equation}
and
\begin{equation}\label{r2u_new}
r_{2}=\frac{B}{\log_{e}2}\sum_{i}\log_{e}\left[1+N\textsc{snr}\lambda_{i}(\Phi_{2})\right]
\end{equation}
where
$\Phi_{1}=\mathbf{H}\mathbf{K}_{x_{1}}\mathbf{H}^{\dag}\mathbf{K}_{z}^{-1}$ and $\Phi_{2}=\mathbf{H}\mathbf{K}_{x_{2}}\mathbf{H}^{\dag}$, and $\lambda_{i}$ is the eigenvalue of the matrices given in the parentheses. Now, we can write the derivatives of $r_{1}$ and $r_{2}$ with respect to $\snr$ as
\begin{equation}\label{r1u_derivative}
\dot{r}_{1}=\frac{B}{\log_{e}2}\sum_{i}\frac{\mu N\lambda_{i}(\Phi_{1})}{1+\mu N\textsc{snr}\lambda_{i}(\Phi_{1})}
\end{equation}
and
\begin{equation}\label{r2u_derivative}
\dot{r}_{2}=\frac{B}{\log_{e}2}\sum_{i}\frac{N\lambda_{i}(\Phi_{2})}{1+N\textsc{snr}\lambda_{i}(\Phi_{2})}.
\end{equation}
Noting that the function $f(\textsc{snr},\theta)$ evaluated at $\textsc{snr}=0$ is 1, i.e., $f(0,\theta)=1$, and $\alpha(0,\theta)$ and $\beta(0,\theta)$ are constants denoted by $\bar{\alpha}$ and $\bar{\beta}$, respectively, we can easily see that the value of the first derivative of the effective rate at $\textsc{snr}=0$ is
\begin{equation}\label{capacity first derivative at snr 0}
\dot{R}_{E}(0,\theta)=\frac{1}{\log_{e}2}\mathbb{E}\left[\bar{\alpha}\mu\tr\{\Phi_{1}\}+\bar{\beta}\tr\{\Phi_{2}\}\right].
\end{equation}

Note that by definition, $\mathbf{K}_{x_{1}}$ and $\mathbf{K}_{x_{2}}$ are positive semi-definite Hermitian matrices. As Hermitian matrices, $\mathbf{K}_{x_{1}}$ and $\mathbf{K}_{x_{2}}$ can be written as follows
\begin{equation}\label{K_Unitary 1}
\mathbf{K}_{x_{1}}=\mathbf{U}_{1}\Lambda_{1}\mathbf{U}_{1}^{\dag}=\sum_{i=1}^{M}\lambda_{1,i}\mathbf{u}_{1,i}\mathbf{u}_{1,i}^{\dag}
\end{equation}
and
\begin{equation}\label{K_Unitary 2}
\mathbf{K}_{x_{2}}=\mathbf{U}_{2}\Lambda_{2}\mathbf{U}_{2}^{\dag}=\sum_{i=1}^{M}\lambda_{2,i}\mathbf{u}_{2,i}\mathbf{u}_{2,i}^{\dag}
\end{equation}
where $\mathbf{U}_{1}$ and $\mathbf{U}_{2}$ are the unitary matrices, $\{\mathbf{u}_{1,i}\}$ and $\{\mathbf{u}_{2,i}\}$ are the column vectors of $\mathbf{U}_{1}$ and $\mathbf{U}_{2}$, respectively. $\Lambda_{1}$ and $\Lambda_{2}$ are the real diagonal matrices with diagonal components $\{\lambda_{1,i}\}$ and $\{\lambda_{2,i}\}$, respectively. Since $\mathbf{K}_{x_{1}}$ and $\mathbf{K}_{x_{2}}$ are positive semi-definite, we have $\lambda_{1,i}\geq0$ and $\lambda_{2,i}\geq0$. Furthermore, since all the available energy should be used for transmission, we have $\tr(\mathbf{K}_{x_{1}})=\sum_{i=1}^{M}\lambda_{1,i}=1$ and $\tr(\mathbf{K}_{x_{2}})=\sum_{i=1}^{M}\lambda_{2,i}=1$.

Now, we can write
\begin{align}
\dot{R}_{E}(0,\theta)&=\frac{1}{\log_{e}2}\mathbb{E}\left[\bar{\alpha}\mu\tr(\mathbf{H}\mathbf{K}_{x_{1}}\mathbf{H}^{\dag}\mathbf{K}_{z}^{-1}) +\bar{\beta}\tr(\mathbf{H}\mathbf{K}_{x_{2}}\mathbf{H}^{\dag})\right]\nonumber\\
&=\frac{1}{\log_{e}2}\mathbb{E}\left[\bar{\alpha}\mu\tr(\mathbf{H}\mathbf{K}_{x_{1}}\mathbf{H}^{\dag}\mathbf{U}_{z}\Lambda_{z}\mathbf{U}_{z}^{\dag}) +\bar{\beta}\tr(\mathbf{H}\mathbf{K}_{x_{2}}\mathbf{H}^{\dag})\right]\nonumber\\
&=\frac{1}{\log_{e}2}\mathbb{E}\left[\bar{\alpha}\mu\tr(\Lambda_{z}^{1/2}\mathbf{U}_{z}^{\dag}\mathbf{H}\mathbf{K}_{x_{1}}\mathbf{H}^{\dag}\mathbf{U}_{z}\Lambda_{z}^{1/2}) +\bar{\beta}\tr(\mathbf{H}\mathbf{K}_{x_{2}}\mathbf{H}^{\dag})\right]\nonumber\\
&=\frac{1}{\log_{e}2}\sum_{i=1}^{M}\bigg\{\lambda_{1,i}\bar{\alpha}\mu\mathbb{E}[\tr(\Lambda_{z}^{1/2}\mathbf{U}_{z}^{\dag}\mathbf{H}\mathbf{u}_{1,i}\mathbf{u}_{1,i}^{\dag}\mathbf{H}^{\dag}\mathbf{U}_{z}\Lambda_{z}^{1/2})] +\lambda_{2,i}\bar{\beta}\mathbb{E}[\tr(\mathbf{H}\mathbf{u}_{2,i}\mathbf{u}_{2,i}^{\dag}\mathbf{H}^{\dag})]\bigg\}\nonumber\\
&=\frac{1}{\log_{e}2}\sum_{i=1}^{M}\bigg\{\lambda_{1,i}\bar{\alpha}\mu\mathbb{E}[\tr(\mathbf{u}_{1,i}^{\dag}\mathbf{H}^{\dag}\mathbf{U}_{z}\Lambda_{z}^{1/2}\Lambda_{z}^{1/2}\mathbf{U}_{z}^{\dag}\mathbf{H}\mathbf{u}_{1,i})]
+\lambda_{2,i}\bar{\beta}\mathbb{E}[\tr(\mathbf{u}_{2,i}^{\dag}\mathbf{H}^{\dag}\mathbf{H}\mathbf{u}_{2,i})]\bigg\}\nonumber\\
&=\frac{1}{\log_{e}2}\sum_{i=1}^{M}\bigg\{\lambda_{1,i}\bar{\alpha}\mu\mathbb{E}[\tr(\mathbf{u}_{1,i}^{\dag}\mathbf{H}^{\dag}\mathbf{K}_{z}^{-1}\mathbf{H}\mathbf{u}_{1,i})]
+\lambda_{2,i}\bar{\beta}\mathbb{E}[\tr(\mathbf{u}_{2,i}^{\dag}\mathbf{H}^{\dag}\mathbf{H}\mathbf{u}_{2,i})]\bigg\}\nonumber\\
&\leq\frac{1}{\log_{e}2}\bigg\{\bar{\alpha}\mu\mathbb{E}\left[\lambda_{\max}(\mathbf{H}^{\dag}\mathbf{K}_{z}^{-1}\mathbf{H})\right] +\bar{\beta}\mathbb{E}\left[\lambda_{\max}(\mathbf{H}^{\dag}\mathbf{H})\right]\bigg\} \label{rate_upperbound_zero_derivative}
\end{align}
where $\lambda_{\max}(\mathbf{H}^{\dag}\mathbf{K}_{z}^{-1}\mathbf{H})$ and $\lambda_{\max}(\mathbf{H}^{\dag}\mathbf{H})$ denote the maximum eigenvalues of the matrices $\mathbf{H}^{\dag}\mathbf{K}_{z}^{-1}\mathbf{H}$ and $\mathbf{H}^{\dag}\mathbf{H}$. The upper bound in ($\ref{rate_upperbound_zero_derivative}$) can be achieved by choosing the normalized input covariance matrices as
\begin{equation}\label{input covariance 1}
\mathbf{K}_{x_{1}}=\mathbf{u}_{1}\mathbf{u}_{1}^{\dag}
\end{equation}
and
\begin{equation}\label{input covariance 2}
\mathbf{K}_{x_{2}}=\mathbf{u}_{2}\mathbf{u}_{2}^{\dag}
\end{equation}
where $\mathbf{u}_{1}$ and $\mathbf{u}_{2}$ are the unit-norm eigenvectors that correspond to the maximum eigenvalues $\lambda_{max}(\mathbf{H}^{\dag}\mathbf{K}_{z}^{-1}\mathbf{H})$ and $\lambda_{max}(\mathbf{H}^{\dag}\mathbf{H})$. This lets us conclude that
\begin{equation}\label{derivative first effective capacity-app}
\dot{C}_{E}(0,\theta)=\frac{1}{\log_{e}2}\bigg\{\bar{\alpha}\mu\mathbb{E}\left[\lambda_{\max}(\mathbf{H}^{\dag}\mathbf{K}_{z}^{-1}\mathbf{H})\right] +\bar{\beta}\mathbb{E}\left[\lambda_{\max}(\mathbf{H}^{\dag}\mathbf{H})\right]\bigg\}.
\end{equation}
Final expression in (\ref{derivative first effective capacity}) is derived by noticing that $\bar{\alpha} = \frac{bP_{d}+aP_{f}}{a+b}$ and $\bar{\beta} = \frac{a(1-P_{f})}{a+b}$, which are obtained by making use of the transition probability expressions in (\ref{p11ler}) and (\ref{prob2}). Note also that we set $\mu = 1$ since $\dot{C}_{E}(0,\theta)$ is achieved in the low-power regime as $\snr$ and hence $P_2$ approach zero, and constraint in (\ref{Power Threshold_new}) is eventually satisfied in this regime for any interference power constraint $P_{int} >0$ regardless of the value of $\mu$. Choosing $\mu =1$ maximizes the first derivative and leads to the smallest value of the minimum energy per bit.

\subsection{Proof of Theorem \ref{theo:second-deriv}} \label{app:second-deriv}

We first note that the upper bound in (\ref{rate_upperbound_zero_derivative}) and hence the first derivative of the effective capacity and the minimum energy per bit is achieved only if the cognitive radio transmits in the maximal-eigenvalue
eigenspaces
of the matrices $\mathbf{H}^{\dag}\mathbf{K}_{z}^{-1}\mathbf{H}$ and $\mathbf{H}^{\dag}\mathbf{H}$. More specifically, input-covariance matrices should be selected as
\begin{equation}\label{covariance_l 1}
\mathbf{K}_{x_{1}}=\sum_{i=1}^{m_{1}}\kappa_{1i}\mathbf{u}_{1,i}\mathbf{u}_{1,i}^{\dag}
\end{equation}
and
\begin{equation}\label{covariance_l 2}
\mathbf{K}_{x_{2}}=\sum_{i=1}^{m_{2}}\kappa_{2i}\mathbf{u}_{2,i}\mathbf{u}_{2,i}^{\dag}
\end{equation}
for some $\kappa_{1i},\kappa_{2i}\in[0,1]$ satisfying $\sum_{i=1}^{m_1}\kappa_{1i}=1$ and $\sum_{i=1}^{m_2}\kappa_{2i}=1$. Above, $m_1$ and $m_2$ denote the multiplicities of the maximum eigenvalues $\lambda_{\max}(\mathbf{H}^{\dag}\mathbf{K}_{z}^{-1}\mathbf{H})$ and $\lambda_{\max}(\mathbf{H}^{\dag}\mathbf{H})$, respectively, and $\{\mathbf{u}_{1,i}\}$ and $\{\mathbf{u}_{2,i}\}$ are the orthonormal eigenvectors that span the maximal-eigenvalue eigenspaces of $\mathbf{H}^{\dag}\mathbf{K}_{z}^{-1}\mathbf{H}$ and $\mathbf{H}^{\dag}\mathbf{H}$, respectively. The above input covariance structure, which is needed to achieve the minimum energy per bit, is consequently required to achieve the second derivative of the effective capacity and hence the wideband slope.

As for the second derivative, we differentiate $\dot{R}_{E}(\textsc{snr},\theta)$ in $(\ref{derivative first 1})$ with respect to $\textsc{snr}$ once more. In order to obtain a closed-form solution, we concentrate on the special case in which $a+b=1$. Now, we obtain
\begin{align}\label{derivative second 1}
\ddot{R}_{E}(\textsc{snr},\theta)&=\frac{D}{\mathbb{E}\left[f(\textsc{snr},\theta)\right]}\mathbb{E}\left[\ddot{f}(\textsc{snr},\theta)\right] -\frac{D}{\mathbb{E}^{2}\left[f(\textsc{snr},\theta)\right]}\mathbb{E}^{2}\left[\dot{f}(\textsc{snr},\theta)\right]
\end{align}
where
\begin{align}\label{f_first_derivative_simple}
\dot{f}(\textsc{snr},\theta)=-\theta T(aP_{f}+bP_{d})\dot{r}_{1}e^{-\theta Tr_{1}} -\theta Ta(1-P_{f})\dot{r}_{2}e^{-\theta Tr_{2}}
\end{align}
and
\begin{align}\label{f second derivative}
\ddot{f}(\textsc{snr},\theta)=&\theta^{2}T^{2}(bP_{d}+aP_{f})\dot{r}_{1}^{2}e^{-\theta Tr_{1}}+\theta^{2}T^{2}a(1-P_{f})\dot{r}_{2}^{2}e^{-\theta Tr_{2}} -\theta T(bP_{d}+aP_{f})\ddot{r}_{1}e^{-\theta Tr_{1}}\nonumber\\&-\theta Ta(1-P_{f})\ddot{r}_{2}e^{-\theta Tr_{2}}.
\end{align}
Now, we can write the second derivatives of $r_{1}$ and $r_{2}$ as
\begin{equation}\label{r1u_second derivative}
\ddot{r}_{1}=-\frac{B}{\log_{e}2}\sum_{i}\frac{\mu^{2}N^{2}\lambda_{i}^{2}(\Phi_{1})}{\left[1+\mu N\textsc{snr}\lambda_{i}(\Phi_{1})\right]^{2}}
\end{equation}
and
\begin{equation}\label{r2u_second derivative}
\ddot{r}_{2}=-\frac{B}{\log_{e}2}\sum_{i}\frac{N^{2}\lambda_{i}^{2}(\Phi_{2})}{\left[1+N\textsc{snr}\lambda_{i}(\Phi_{2})\right]^{2}}.
\end{equation}

We can easily see that when $\textsc{snr}$ goes to 0, we can express the first and second derivatives of $f(\textsc{snr},\theta)$
\begin{equation}\label{f_first_derivative}
\dot{f}(0,\theta)=-\frac{(bP_{d}+aP_{f})\theta TBN\mu}{\log_{e}2}\tr\{\Phi_{1}\}-\frac{a(1-P_{f})\theta TBN}{\log_{e}2}\tr\{\Phi_{2}\}
\end{equation}
and
\begin{align}\label{f_second_derivative}
\ddot{f}(0,\theta)&=\frac{\ell_{1}\theta TBN^{2}\mu^{2}}{\log_{e}2}\tr\{\Phi_{1}^{\dag}\Phi_{1}\} + \frac{\ell_{2}\theta TBN^{2}}{\log_{e}2}\tr\{\Phi_{2}^{\dag}\Phi_{2}\}\nonumber\\ &+ \frac{\ell_{1}\theta^{2}T^{2}B^{2}N^{2}\mu^{2}}{\log_{e}^{2}2}\tr^{2}\{\Phi_{1}\} + \frac{\ell_{2}\theta^{2}T^{2}B^{2}N^{2}}{\log_{e}^{2}2}\tr^{2}\{\Phi_{2}\},
\end{align}
and $\ell_{1}=(bP_{d}+aP_{f})$ and $\ell_{2}=a(1-P_{f})$.
We know $f(0,\theta)=1$. Then, we write
\begin{equation}
\ddot{R}(0,\theta)=\frac{1}{\theta TBN}\left\{\mathbb{E}^{2}\left[\dot{f}(0,\theta)\right]-\mathbb{E}\left[\ddot{f}(0,\theta)\right]\right\}.
\end{equation}

We can easily verify that
\begin{align}\label{verification 1}
\mathbb{E}\left\{\tr(\Phi_{1})\right\}&=\mathbb{E}\left\{\tr(\mathbf{H}\mathbf{K}_{x_{1}}\mathbf{H}^{\dag}\mathbf{K}_{z}^{-1})\right\} =\mathbb{E}\left\{\lambda_{\max}(\mathbf{H}^{\dag}\mathbf{K}_{z}^{-1}\mathbf{H})\right\}\\
\mathbb{E}\left\{\tr(\Phi_{2})\right\}&=\mathbb{E}\left\{\tr(\mathbf{H}\mathbf{K}_{x_{2}}\mathbf{H}^{\dag})\right\}
=\mathbb{E}\left\{\lambda_{\max}(\mathbf{H}^{\dag}\mathbf{H})\right\}
\end{align}
and
\begin{align}\label{verification 2}
\mathbb{E}\left\{\tr(\Phi_{1}^{\dag}\Phi_{1})\right\}&=\mathbb{E}\left\{\tr(\mathbf{K}_{z}^{-1}\mathbf{H}\mathbf{K}_{x_{1}}\mathbf{H}^{\dag} \mathbf{H}\mathbf{K}_{x_{1}}\mathbf{H}^{\dag}\mathbf{K}_{z}^{-1} )\right\}\nonumber\\&=\mathbb{E}\left\{\tr(\mathbf{K}_{z}^{-1}\mathbf{K}_{z}^{-1}\mathbf{H}\mathbf{K}_{x_{1}}\mathbf{H}^{\dag} \mathbf{H}\mathbf{K}_{x_{1}}\mathbf{H}^{\dag})\right\}\\
&\geq\mathbb{E}\left\{\tr(\mathbf{K}_{z}^{-1}\mathbf{H}\mathbf{K}_{x_{1}}\mathbf{H}^{\dag} \mathbf{K}_{z}^{-1}\mathbf{H}\mathbf{K}_{x_{1}}\mathbf{H}^{\dag})\right\}\label{kitaptan}\\
&=\mathbb{E}\left\{\sum_{i,j}^{m_1}\kappa_{1i}\kappa_{1j}\tr(\mathbf{K}_{z}^{-1}\mathbf{H}\mathbf{u}_{i}\mathbf{u}_{i}^{\dag}\mathbf{H}^{\dag} \mathbf{K}_{z}^{-1}\mathbf{H}\mathbf{u}_{j}\mathbf{u}_{j}^{\dag}\mathbf{H}^{\dag})\right\}\label{follow1}\\
&=\mathbb{E}\left\{\sum_{i}^{m_1}\kappa_{1i}^{2}\tr(\mathbf{K}_{z}^{-1}\mathbf{H}\mathbf{u}_{i}\mathbf{u}_{i}^{\dag}\mathbf{H}^{\dag} \mathbf{K}_{z}^{-1}\mathbf{H}\mathbf{u}_{i}\mathbf{u}_{i}^{\dag}\mathbf{H}^{\dag})\right\}\label{follow2}\\
&=\mathbb{E}\left\{\sum_{i}^{m_1}\kappa_{1i}^{2}\lambda_{\max}(\mathbf{H}^{\dag} \mathbf{K}_{z}^{-1}\mathbf{H})\tr(\mathbf{K}_{z}^{-1}\mathbf{H}\mathbf{u}_{i}\mathbf{u}_{i}^{\dag}\mathbf{H}^{\dag})\right\}\\
&=\mathbb{E}\left\{\sum_{i}^{m_1}\kappa_{1i}^{2}\lambda_{\max}(\mathbf{H}^{\dag} \mathbf{K}_{z}^{-1}\mathbf{H})\tr(\mathbf{u}_{i}^{\dag}\mathbf{H}^{\dag}\mathbf{K}_{z}^{-1}\mathbf{H}\mathbf{u}_{i})\right\}\\
&=\mathbb{E}\left\{\sum_{i}^{m_1}\kappa_{1i}^{2}\lambda_{\max}^{2}(\mathbf{H}^{\dag} \mathbf{K}_{z}^{-1}\mathbf{H})\right\}\\
&=\mathbb{E}\left\{\lambda_{\max}^{2}(\mathbf{H}^{\dag} \mathbf{K}_{z}^{-1}\mathbf{H})\sum_{i}^{m_1}\kappa_{1i}^{2}\right\}\\
&\geq\frac{1}{m_1}\mathbb{E}\left\{\lambda_{\max}^{2}(\mathbf{H}^{\dag}\mathbf{K}_{z}^{-1}\mathbf{H})\right\}\label{follow3}
\end{align}
where ($\ref{kitaptan}$) comes from the fact that if $A,B\in M_{n}$ are Hermitian, $\tr(AB)^{2}\leq\tr(A^{2}B^{2})$ \cite[Chap. 4, Problem 4.1.11]{Matrix Analysis}. ($\ref{follow1}$) and ($\ref{follow2}$) follow from the fact that $\{\mathbf{u}_{1i}\}$ are the eigenvectors that correspond to $\lambda_{\max}(\mathbf{H}^{\dag}\mathbf{K}_{z}^{-1}\mathbf{H})$ and hence $\mathbf{u}_{1,i}^{\dag}\mathbf{H}^{\dag}\mathbf{K}_{z}^{-1}\mathbf{H}\mathbf{u}_{1,j}=0$ if $i\neq j$, which comes from the orthonormality of $\{\mathbf{u}_{1,i}\}$. Finally, ($\ref{follow3}$) follows from the properties that $\kappa_{1i}\in[0,1]$ and $\sum_{i=1}^{m_1}\kappa_{1i}=1$, and the fact that $\sum_{i=1}^{m_1}\kappa_{1i}^{2}$ is minimized by choosing $\kappa_{1i}=\frac{1}{m_1}$, that leads us to the lower bound $\sum_{i=1}^{m_1}\kappa_{1i}^{2}\geq\frac{1}{m_1}$. Same procedure can be applied to $\mathbb{E}\left\{\tr(\Phi_{2}^{\dag}\Phi_{2})\right\}$, and we can easily see that

\begin{align}\label{verification 2_1}
\mathbb{E}\left\{\tr(\Phi_{2}^{\dag}\Phi_{2})\right\}=\mathbb{E}\left\{\tr(\mathbf{H}\mathbf{K}_{x_{1}}\mathbf{H}^{\dag} \mathbf{H}\mathbf{K}_{x_{1}}\mathbf{H}^{\dag})\right\}&=\mathbb{E}\left\{\sum_{i,j}^{m_2}\kappa_{2,i}\kappa_{2,j} \tr(\mathbf{H}\mathbf{u}_{2,i}\mathbf{u}_{2,i}^{\dag}\mathbf{H}^{\dag}\mathbf{H}\mathbf{u}_{2,j}\mathbf{u}_{2,j}^{\dag} \mathbf{H}^{\dag})\right\}\\
&=\mathbb{E}\left\{\sum_{i}^{m_2}\kappa_{2,i}^{2} \tr(\mathbf{H}\mathbf{u}_{2,i}\mathbf{u}_{2,i}^{\dag}\mathbf{H}^{\dag}\mathbf{H}\mathbf{u}_{2,i}\mathbf{u}_{2,i}^{\dag} \mathbf{H}^{\dag})\right\}\\
&=\mathbb{E}\left\{\sum_{i}^{m_2}\kappa_{2,i}^{2}\lambda_{\max}(\mathbf{H}^{\dag}\mathbf{H}) \tr(\mathbf{H}\mathbf{u}_{2,i}\mathbf{u}_{2,i}^{\dag}\mathbf{H}^{\dag})\right\}\\
&=\mathbb{E}\left\{\sum_{i}^{m_2}\kappa_{2,i}^{2}\lambda_{\max}(\mathbf{H}^{\dag}\mathbf{H}) \tr(\mathbf{u}_{2,i}^{\dag}\mathbf{H}^{\dag}\mathbf{H}\mathbf{u}_{2,i})\right\}\\
&=\mathbb{E}\left\{\sum_{i}^{m_2}\kappa_{2,i}^{2}\lambda_{\max}^{2}(\mathbf{H}^{\dag}\mathbf{H})\right\}\\
&=\mathbb{E}\left\{\lambda_{\max}^{2}(\mathbf{H}^{\dag}\mathbf{H})\sum_{i}^{m_2}\kappa_{2,i}^{2}\right\}\\
&\geq\frac{1}{m_2}\mathbb{E}\left\{\lambda_{\max}^{2}(\mathbf{H}^{\dag}\mathbf{H})\right\}
\end{align}

Now, we can write the second derivative of effective rate as
\begin{align}\label{second derivative of effective rate}
\ddot{R}_{E}(0,\theta)=&\frac{1}{\theta TBN}\bigg\{\mathbb{E}^{2}\left[\frac{\ell_{1}\theta TBN\mu}{\log_{e}2}\tr(\Phi_{1})+\frac{\ell_{2}\theta TBN}{\log_{e}2}\tr(\Phi_{2})\right]-\mathbb{E}\bigg[ \frac{\ell_{1}\theta^{2}T^{2}B^{2}N^{2}\mu^{2}}{\log_{e}^{2}2}\tr^{2}(\Phi_{1})\nonumber\\ &+ \frac{\ell_{2}\theta^{2}T^{2}B^{2}N^{2}}{\log_{e}^{2}2}\tr^{2}(\Phi_{2})\bigg] -\mathbb{E}\bigg[\frac{\ell_{1}\theta TBN^{2}\mu^{2}}{\log_{e}2}\tr(\Phi_{1}^{\dag}\Phi_{1})+ \frac{\ell_{2}\theta TBN^{2}}{\log_{e}2}\tr(\Phi_{2}^{\dag}\Phi_{2}) \bigg]\bigg\}\\
=&\frac{\theta TBN}{\log_{e}^{2}2}\mathbb{E}^{2}\left[\ell_{1}\mu\tr(\Phi_{1})+\ell_{2}\tr(\Phi_{2})\right]-\frac{\theta TBN}{\log_{e}^{2}2}\mathbb{E}\left[\ell_{1}\mu^{2}\tr^{2}(\Phi_{1})+\ell_{2}\tr^{2}(\Phi_{2})\right]\nonumber\\
&-\frac{N}{\log_{e}2}\mathbb{E}\left[\ell_{1}\mu^{2}\tr(\Phi_{1}^{\dag}\Phi_{1})+\ell_{2}\tr(\Phi_{2}^{\dag}\Phi_{2})\right]\\
\leq&\frac{\theta TBN}{\log_{e}^{2}2}\mathbb{E}^{2}\left[\ell_{1}\mu\lambda_{\max}(\mathbf{H}^{\dag}\mathbf{K}_{z}^{-1}\mathbf{H})+\ell_{2} \lambda_{\max}(\mathbf{H}^{\dag}\mathbf{H})\right]\nonumber\\&-\frac{\theta TBN}{\log_{e}^{2}2}\mathbb{E}\left[\ell_{1}\mu^{2}\lambda_{\max}^{2}(\mathbf{H}^{\dag}\mathbf{K}_{z}^{-1}\mathbf{H}) +\ell_{2}\lambda_{\max}^{2}(\mathbf{H}^{\dag}\mathbf{H})\right]\nonumber\\
&-\frac{N}{\log_{e}2}\mathbb{E}\left[\frac{\ell_{1}\mu^{2}\lambda_{\max}^{2}(\mathbf{H}^{\dag}\mathbf{K}_{z}^{-1}\mathbf{H})}{m_1} +\frac{\ell_{2}\lambda_{\max}^{2}(\mathbf{H}^{\dag}\mathbf{H})}{m_2}\right]\nonumber\\=&\ddot{C}_{E}(0,\theta)\label{second derivative of effective rate and capacity}
\end{align}

Finally, we again set $\mu =1$ following the same reasoning discussed at the end of Appendix \ref{app:first-deriv}.

\end{spacing}

\end{document}